\documentclass[12pt]{article}
\usepackage{amsmath}
\usepackage{latexsym}
\usepackage{mathrsfs}
\usepackage{amsfonts} 
\usepackage{amssymb}  
\usepackage{array}
\usepackage{bm}

\usepackage{colortbl}
\usepackage{multirow}
\usepackage{graphicx}
\definecolor{gray80}{gray}{0.8}

\DeclareSymbolFont{boldletters}{OML}{cmm} {b}{it}
\DeclareSymbolFontAlphabet{\mathbit}{boldletters}
\DeclareMathSymbol{\alpha}{\mathalpha}{letters}{"0B}
\DeclareMathSymbol{\beta}{\mathalpha}{letters}{"0C}
\DeclareMathSymbol{\gamma}{\mathalpha}{letters}{"0D}
\DeclareMathSymbol{\delta}{\mathalpha}{letters}{"0E}
\DeclareMathSymbol{\epsilon}{\mathalpha}{letters}{"0F}
\DeclareMathSymbol{\zeta}{\mathalpha}{letters}{"10}
\DeclareMathSymbol{\eta}{\mathalpha}{letters}{"11}
\DeclareMathSymbol{\theta}{\mathalpha}{letters}{"12}
\DeclareMathSymbol{\iota}{\mathalpha}{letters}{"13}
\DeclareMathSymbol{\kappa}{\mathalpha}{letters}{"14}
\DeclareMathSymbol{\lambda}{\mathalpha}{letters}{"15}
\DeclareMathSymbol{\mu}{\mathalpha}{letters}{"16}
\DeclareMathSymbol{\nu}{\mathalpha}{letters}{"17}
\DeclareMathSymbol{\xi}{\mathalpha}{letters}{"18}
\DeclareMathSymbol{\pi}{\mathalpha}{letters}{"19}
\DeclareMathSymbol{\rho}{\mathalpha}{letters}{"1A}
\DeclareMathSymbol{\sigma}{\mathalpha}{letters}{"1B}
\DeclareMathSymbol{\tau}{\mathalpha}{letters}{"1C}
\DeclareMathSymbol{\upsilon}{\mathalpha}{letters}{"1D}
\DeclareMathSymbol{\phi}{\mathalpha}{letters}{"1E}
\DeclareMathSymbol{\chi}{\mathalpha}{letters}{"1F}
\DeclareMathSymbol{\psi}{\mathalpha}{letters}{"20}
\DeclareMathSymbol{\omega}{\mathalpha}{letters}{"21}
\DeclareMathSymbol{\varepsilon}{\mathalpha}{letters}{"22}
\DeclareMathSymbol{\vartheta}{\mathalpha}{letters}{"23}
\DeclareMathSymbol{\varpi}{\mathalpha}{letters}{"24}
\DeclareMathSymbol{\varrho}{\mathalpha}{letters}{"25}
\DeclareMathSymbol{\varsigma}{\mathalpha}{letters}{"26}
\DeclareMathSymbol{\varphi }{\mathalpha}{letters}{"27}
\DeclareMathSymbol{\Gamma}{\mathalpha}{letters}{"00}
\DeclareMathSymbol{\Delta}{\mathalpha}{letters}{"01}
\DeclareMathSymbol{\Theta}{\mathalpha}{letters}{"02}
\DeclareMathSymbol{\Lambda}{\mathalpha}{letters}{"03}
\DeclareMathSymbol{\Xi}{\mathalpha}{letters}{"04}
\DeclareMathSymbol{\Pi}{\mathalpha}{letters}{"05}
\DeclareMathSymbol{\Sigma}{\mathalpha}{letters}{"06}
\DeclareMathSymbol{\Upsilon}{\mathalpha}{letters}{"07}
\DeclareMathSymbol{\Phi}{\mathalpha}{letters}{"08}
\DeclareMathSymbol{\Psi}{\mathalpha}{letters}{"09}
\DeclareMathSymbol{\Omega}{\mathalpha}{letters}{"0A}
\newcommand{\mbit}[1]{{\mathbit#1}}

\newcommand{\Nabla}{{\mbox{\boldmath$\nabla$}}}
\newcommand{\dbox}{\,\framebox(7,7)[t]{}\,}
\newcommand{\dsl}[2]{{#1}{\mbox{\hspace{-8pt}\hspace{#2}$\not$}
\hspace{8pt}\hspace{-#2}}} 

\newcommand{\dslpar}{\dsl{\partial}{0pt}}
\title{Relativistic Remnants of Non-relativistic Electrons}
\author{Taro Kashiwa\thanks{kashiwa@phys.sci.ehime-u.ac.jp} \ and  Taisuke Yamaguchi\thanks{yama2607@gmail.com} \\
Department of Physics, Graduate School of Science and Engineering,\\
 Ehime University, Matsuyama 790-8577, Japan \\
\\
}
\date{\today}
\begin{document}
\maketitle

\abstract{Electrons obeying the Dirac equation are investigated under the non-relativistic $c \mapsto \infty$ limit.  General solutions are given by derivatives of relativistic invariant functions which possess discontinuity at the light-cone, yielding the delta function of $(ct)^2 - \mbit{x}^2$. This light-cone singularity does survive in this limit to show that the charge and the current densities of electrons travel with the speed of light in spite of their massiveness.}

\section{Introduction}\label{Introduction}

It is well known that the Dirac electron has a piece vibrating with the light velocity, called "Zitterbewegung" whose origin is supposed as a mixing of the positive and the negative components. Discussions have been made by the Heisenberg equation of motion for the Dirac Hamiltonian\cite{rf:Zitter1} or by the momentum representation for the Dirac field $\Psi(x)$\cite{rf:Zitter2}. Electron moving in the zigzag motion with the light speed also appears on the stage of the pilot-wave approach to quantum field theory\cite{rf:CW} (Feynman had already discussed the zigzag motion with the light velocity in the context of path integral\cite{rf:FeyHibbs}).

We shall, in this paper, focus the electron field $\Psi(x)$ itself, to show that electrons obeying the Dirac equation inevitably bear portions traveling with the light speed in the non-relativistic limit. As a preliminary, let us recall the Dirac Hamiltonian
\begin{eqnarray}
H_{\rm D} = c \left( \mbit{ \alpha} \cdot \mbit{p} + \beta mc \right)   \  ;  \qquad \mbit{\alpha} \equiv \gamma^0 \mbit{\gamma} \ ;  \  \beta \equiv \gamma^0  \   ,   \label{DiracHamiltonian}   
\end{eqnarray}
with
\begin{eqnarray}
\left\{ \gamma^\mu , \gamma^\nu \right\} = 2 \eta^{\mu \nu}   \ ,   \quad  {\rm diag}( \eta^{\mu \nu} ) = (1, -1,-1,-1)   \  ;  \ (\mu, \nu = 0,1,2,3) \ ,  
\end{eqnarray}
being the $4\times4$ gamma matrices represented as
\begin{eqnarray}
\gamma^0  =\left(\begin{array}{cc} {\bf I} & 0 \\
0 & - {\bf I} 
\end{array}\right)  \  ,  \  \mbit{ \gamma}   =\left(\begin{array}{cc} 0 & \mbit{\sigma}  \\
- \mbit{\sigma}   & 0   
\end{array}\right)  \  ;  \quad  \mbox{ $\mbit{\sigma}$: Pauli matrices}   \  .   \label{gammaMatrix}
\end{eqnarray}
In order to investigate the non-relativistic limit, it is useful to perform a unitary (called the Foldy-Wouthuysen) transformation\cite{rf:BD},
\begin{eqnarray}
 U \equiv \exp \! \left[\frac{ \mbit{\gamma} \cdot \mbit{p} }{mc}  \theta(\mbit{p})  \right] = \cos \left( \frac{| \mbit{p} |}{mc}  \theta(\mbit{p})     \right) + \frac{ \mbit{\gamma} \cdot \mbit{p} }{ | \mbit{p} | } \sin \left( \frac{| \mbit{p} |}{mc}  \theta(\mbit{p})     \right)   \ ,    \label{FWTransformation}
\end{eqnarray}
with
$$
\tan \left(  2 \frac{|\mbit{p} | }{mc} \theta(\mbit{p}) \right) =   \frac{ |\mbit{p}| }{mc}  \ , 
$$
such that
\begin{eqnarray}
 UH_{\rm D} U^\dagger = \beta H  \   ;   \qquad      H \equiv  c \sqrt{ \mbit{p}^2 + m^2 c^2 }  \ .     \label{3DHamiltonian}
\end{eqnarray}
Here the lower component corresponds to the negative energy state which should be discarded in the non-relativistic world. Therefore we shall pick up the positive energy part and study its wave mechanical structure in the next section. 
In the following section \ref{Sec:The Charge Density of Free Electrons}, we shall treat covariant solutions of the free Dirac equation,  which will be extended to interacting cases in sec.\ref{Sec:The Charge and the Current Density for Electrons in a Laboratory}. The final section is devoted to the discussion. Some of the detailed calculations in sec.\ref{Sec:The Charge and the Current Density for Electrons in a Laboratory} are relegated to the appendices.

\section{Wave Mechanics of $H= c \sqrt{ \mbit{p}^2 + m^2 c^2 } $}\label{Sec:ToyModel}

Consider a single component wave mechanics governed by the Hamiltonian (\ref{3DHamiltonian}), that is, a wave function $\Psi(t, \mbit{x} )$ obeying the Schr\"{o}dinger equation
\begin{eqnarray}
i \hbar \frac{\partial }{\partial t} \Psi(t, \mbit{x} ) =  c \sqrt{ - (\hbar \Nabla) ^2 + m^2 c^2 } \ \Psi(t, \mbit{x} )  \  .  
\end{eqnarray}
The solution reads
\begin{eqnarray}
\Psi(t, \mbit{x} ) =  \int d^3 \mbit{x}'  K\! \left( \Delta \mbit{x} ; t\right) \psi (0, \mbit{x}')       \  ,     \label{Psi x}
\end{eqnarray}
where $\psi(0, \mbit{x}')$ is an arbitrary function,
\begin{eqnarray}
& & \hspace{-5ex} K \! \left( \Delta \mbit{x}  ;  t \right) \equiv \langle \mbit{x} | \exp \!  \left[  - i  \frac{c t  }{\hbar} \sqrt{ \hat{\mbit{P} }^2 + m^2 c^2 } \ \right] | \mbit{x}' \rangle \nonumber \\
& &  = \int \frac{d ^3 \mbit{p} }{ (2  \pi \hbar)^3 } \exp \left[\frac{ i }{\hbar} \left( \mbit{p} \cdot \Delta \mbit{x}  - c t \sqrt{ \mbit{p}^2 + m^2 c^2 } \right) \right]   \    ;   \quad  \Delta \mbit{x} \equiv \mbit{x} - \mbit{x}'   \  ,    \label{kernel}
\end{eqnarray}
is the kernel\footnote{In the path integral formalism, this is a typical example that the Hamiltonian prescription is more general\cite{rf:KOSGarrod} than Feynman's\cite{rf:Feynman}, whose Euclidean case, that is, $t \mapsto -it$ in (\ref{kernel}) has been discussed in \cite{rf:FKashiwa}. } with $\hat{\mbit{P}}$ designating the momentum operator and the integration range from $-\infty$ to $\infty$ has been omitted here and hereafter unless otherwise specified.

\vspace{1ex}

In order to calculate the kernel (\ref{kernel}), introduce $\mbit{k} \equiv \mbit{p}/ \hbar$ and write
\begin{eqnarray}
\mu \equiv \frac{ mc}{\hbar}   \  ,  \label{Defmu}
\end{eqnarray}
to obtain
\begin{eqnarray}
 K \! \left( \Delta \mbit{x}  ; t\right) = \int \frac{ d^3 \mbit{k} }{ (2 \pi)^3} \exp \! \left[ i \mbit{k} \cdot \Delta \mbit{x} - i c t\sqrt{\mbit{k}^2 + \mu^2 } \right]=    - \frac{1}{4 \pi^2 r}  \frac{\partial I (r, t) }{\partial r}   \  ;     \label{kenelMomentumRep}   \\
 I( r , t)  \equiv \int_{-\infty }^{\infty } dk \exp\left[  i \left(  kr  -c t\sqrt{k^2 + \mu^2 }  \right)   \right]  \  ;  \qquad    r \equiv | \Delta \mbit{x}  |   \  ,   \hspace{5ex}   \label{DefI}
\end{eqnarray}
where use has been made of the polar coordinates to the final expression. Put $k = \mu \sinh \Theta$ to find
\begin{eqnarray}
I( r , t) = \mu  \int_{-\infty }^{\infty } d \Theta \cosh \Theta \exp \! \left[ -i\mu \left( c t\cosh \Theta - r \sinh \Theta  \right)  \right]  \ ,
\end{eqnarray}
whose exponent reads, with the aid of an addition theorem, as
\begin{eqnarray}
ct \cosh \Theta - r \sinh \Theta = \left\{
\begin{array}{ccc}
 \hspace{-1ex}  \sqrt{x_\mu^2} \cosh (\Theta -\alpha )    &  \hspace{-2ex} : 
    \  \displaystyle{\tanh \alpha = \frac{r }{ ct }   }  \  ; 
    &  \hspace{-1ex} ct   >  r      \\
\noalign{\vspace{2ex}}
\hspace{-1ex} - \sqrt{-x_\mu^2} \sinh (\Theta -\beta )    & \hspace{-2ex}  :
  \  \displaystyle{ \tanh \beta  =  \frac{ct}{ r }   }   \ ;   &   \hspace{-1ex}  r > ct    
\end{array}           \right.  \  ,    \label{tanh alpha,beta}
\end{eqnarray}
with 
\begin{eqnarray}
 x_\mu^2\equiv (c t)^2 - r^2   \ .  \label{x^2 Definition}
\end{eqnarray}
Make sifts $\Theta \mapsto \Theta + \alpha, \beta$ and again utilize the addition theorem to obtain
\begin{eqnarray}
& & \hspace{-8ex} I( r , t) =  \mu \theta (ct -r) \cosh \alpha  \int_{-\infty }^{\infty } \hspace{-1ex} d \Theta    \cosh \Theta  {\mathrm e}^{-i \chi_{+} \cosh \Theta  }   \nonumber  \\
& & \hspace{-4ex} 
+  \mu  \theta (r-ct)   \left[    \cosh \beta   \int_{-\infty }^{\infty }  \hspace{-1ex} d \Theta \cosh \Theta   {\mathrm e}^{i \chi_{-} \sinh \Theta  }  +  \sinh \beta   \int_{-\infty }^{\infty }  \hspace{-2ex} d \Theta \sinh \Theta   {\mathrm e}^{i \chi_{-} \sinh \Theta  }   \right]  , \label{I-1}  
\end{eqnarray}
with
\begin{eqnarray}
\chi_{\pm } \equiv \mu \sqrt{\pm x_\mu^2 }    \  .    \label{chi+-Definition}
\end{eqnarray}
(Here we have discarded the odd function part:$\displaystyle{\int_{-\infty }^{\infty } \hspace{-2ex} d \Theta  \sinh \Theta  {\mathrm e}^{-i \chi_{+} \cosh \Theta  } =0}$.) By noting that
\begin{eqnarray*}
\int_{-\infty }^{\infty }  \hspace{-1ex} d \Theta \cosh \Theta   {\mathrm e}^{i \chi_{-} \sinh \Theta  }  = \int_{-\infty }^{\infty }  \hspace{-1ex} d (\sinh \Theta )    {\mathrm e}^{i \chi_{-} \sinh \Theta  }  = 2 \pi \delta(\chi_{-} ) = \frac{2 \pi}{\mu} \delta (\sqrt{-x_\mu^2} )     \  ,
\end{eqnarray*}
and $\cosh \beta = r/ \sqrt{- x_\mu^2}$ in view of (\ref{tanh alpha,beta}), the second term of (\ref{I-1}) reads
\begin{eqnarray}
 \mu  \theta (r-ct) \cosh \beta \int_{-\infty }^{\infty }  \hspace{-1ex} d \Theta \cosh \Theta   {\mathrm e}^{i \chi_{-} \sinh \Theta  }= \theta (r -ct)\left(  \mu \frac{r}{\sqrt{-x_\mu^2}}  \right) \left(  \frac{2 \pi}{\mu} \delta (\sqrt{-x_\mu^2} ) \right)   \nonumber  \\
=  4 \pi r \theta (r -ct) \delta (x^2_\mu )  =\theta (r -ct)  \frac{2 \pi r}{ct}\delta (ct-r) = \pi \delta(ct-r)    \  ,     \hspace{8ex}  
\end{eqnarray}
where we have used the relations, 
$$
\delta (\sqrt{-x_\mu^2} ) = 2 \sqrt{- x_\mu^2}  \delta (x_\mu^2)  \ ,   \qquad  \theta(0)= \frac{1}{2}   \  . 
$$
Finally the Bessel function formulas\cite{rf:Ryz}
\begin{eqnarray*}
 \int_{-\infty }^{\infty } \hspace{-1ex} d \Theta    \cosh \Theta \  {\mathrm e}^{-i \chi_{+} \cosh \Theta  }  = - \pi  H_1^{(2)} \! \left( \chi_{+} \right)  \ ,            \\
 \int_{-\infty }^{\infty }  \hspace{-1ex} d \Theta \sinh \Theta  \  {\mathrm e}^{i \chi_{-} \sinh \Theta  }  = 2i K_1\! \left( \chi_{-} \right)  \ ,  \hspace{3ex}    
\end{eqnarray*}
lead us to
\begin{eqnarray}
I( r , t)  =  \pi  \delta (c t-r)   - 2 ct\mu^2 \left[  \theta (ct-r ) \left(  \frac{\pi }{2}   \frac{  H_1^{(2)} \! \left( \chi_{+} \right) }{\chi_{+}} \right)   -  \theta (r -c t)\left( i \frac{ K_1 \! \left( \chi_{-} \right) }{\chi_{-}} \right)  \right]   \    .      \label{I Result}
\end{eqnarray}
Here note that $I$ consists of different functions in the regions $ct > r$ and $ct < r$, which causes the delta function singularity when a differentiation is made. (The $\mu$-independent delta function emerges from the huge momentum domain $\mbit{p} \gg mc$.)

The kernel is, from (\ref{kenelMomentumRep}), obtained, by differentiating (\ref{I Result}) with respect to $r$, as 
\begin{eqnarray}
& &  \hspace{-6ex} K\! \left( \Delta \mbit{x} ; t\right)
= - \frac{1 }{4 \pi r }\frac{\partial }{\partial r} \delta \! \left( c t- r  \right) - \frac{i \mu^2 }{ 4 \pi^2 }\delta \! \left( c t-r  \right)   \nonumber \\
& & \hspace{0ex}  + \frac{ct \mu^2}{2 \pi^2}\left[ \theta (ct-r) \frac{ \pi }{ 2  } \frac{ H_2^{(2)} \! \left( \mu \sqrt{x_\mu^2}  \right)}{x_\mu^2}  - i \theta (r -c t)  \frac{K_2 \! \left( \mu \sqrt{-x_\mu^2} \right)}{ x_\mu^2  } \right]   \   ,  \label{3DkernelResult}
\end{eqnarray}
where use has been made of 
\begin{eqnarray*}
 \frac{\partial }{\partial r}\theta (ct-r ) = - \delta (ct -r)  \ ;   \quad  \frac{\partial }{\partial r}\theta (r -ct)  =  \delta (ct -r)   \  ,  
\end{eqnarray*}
and then (see Appendix \ref{AppendixA})
\begin{eqnarray}
-\frac{ct\mu^2}{2 \pi^2 r } \delta \! \left( ct-r  \right) \left[ \frac{\pi }{2} \frac{   H_1^{(2)} \! \left( \chi_{+} \right) }{\chi_{+}} + i  \frac{  K_1 \! \left( \chi_{-} \right) }{\chi_{-}}  \right] =  - \frac{i \mu^2 }{ 4 \pi^2 }\delta \! \left( c t-r  \right)    \ , \label{H^(2)-K zero} 
\end{eqnarray}
as well as \cite{rf:Ryz3},
\begin{eqnarray}
\frac{d}{z dz} \left( \frac{ Z_1 (z ) }{ z} \right) =  - \frac{ Z_2 (z ) }{ z^2 }  \   ,    \quad  \mbox{$Z_n \  :  \  J_n , N_n , ( \mbox{also $H^{(2)}_n$}) $ and $K_n$ }        \  ,    \label{BesselDifferntiation}
\end{eqnarray}
by noting
\begin{eqnarray}
\frac{1}{r}\frac{\partial }{\partial r}  \stackrel{(\ref{chi+-Definition})}=  \mp \mu^2 \frac{1}{\chi_{\pm }}\frac{\partial }{\partial \chi_{\pm }}  \   .  
\end{eqnarray}

By taking the non-relativistic limit $c \mapsto \infty$, that is, $\mu \mapsto \infty$ (\ref{Defmu}), the kernel(\ref{3DkernelResult}) reads, 
\begin{eqnarray}
K\! \left( \Delta \mbit{x} ; t\right) \stackrel{\mu \mapsto \infty}{=}  - \frac{i \mu^2 }{ 4 \pi^2 }\delta \! \left( c t-| \Delta \mbit{x} |  \right) +  O \! \left( \mu^{3/2} \right)   \  ,     \label{Kernel mu Large}
\end{eqnarray}
where we have employed the asymptotic expansion of the Bessel function\cite{rf:Ryz4}
\begin{eqnarray}
Z_n(z) \stackrel{z \mapsto \infty}= O \! \left( z^{- 1/2} \right) , \  Z_n : J_n , N_n , \  (  \mbox{also $H^{(2)}_n$}) \ ;  \quad K_n(z)  \stackrel{z \mapsto \infty}= \frac{{\mathrm e}^{-z} }{\sqrt{z} }(1 + O \! \left( z^{-2} \right) )     \   .     \label{asymExp Bessel}
\end{eqnarray}

Now put
$$
 \psi(0, \mbit{x}') = \left( \frac{1}{a \pi} \right)^{3/4} \exp \! \left(  -  \frac{ {\mbit{x}' }^2 }{ 2 a} \right)   \  , 
$$
then substitute (\ref{Kernel mu Large}) into (\ref{Psi x}) with changing the variables as $\mbit{x}' \mapsto \Delta \mbit{x}$ to find
\begin{eqnarray}
 \Psi(t, \mbit{x} ) = -  \frac{i \mu^2}{4 \pi} \left( \frac{1}{a \pi} \right)^{3/4}  \int d^3 (\Delta \mbit{x})  \delta \left( c t  - | \Delta \mbit{x} | \right) \exp \! \left[  -  \frac{ \left( \mbit{x} + \Delta \mbit{x}   \right)^2 }{2a}  \right]  +   O \! \left( \mu^{3/2} \right)  \  , 
\end{eqnarray}
which, by introducing the polar coordinates, yields
\begin{eqnarray}
 \Psi(t, \mbit{x} ) = - \frac{ i \mu^2 a^{1/4} }{ \pi^{3/4} } \frac{ct }{ |\mbit{x}| } \left[  \exp \! \left(- \frac{ (ct - |\mbit{x}| )^2 }{2 a}     \right)  -   \exp \! \left(- \frac{ (ct + |\mbit{x}| )^2 }{2 a}     \right)     \right]     \  .     \label{WaveFn Result}
\end{eqnarray}
When $ ct, |\mbit{x} | \gg \sqrt{a} $, the second term fades away and around the peak,
\begin{eqnarray}
ct - |\mbit{x} | \approx 0    \  ,   \label{The Peak} 
\end{eqnarray}
(\ref{WaveFn Result}) becomes
\begin{eqnarray}
\Psi(t, \mbit{x} ) \approx   - \frac{ i \mu^2 a^{1/4} }{ \pi^{3/4} } \exp \! \left(- \frac{ (ct - |\mbit{x}| )^2 }{2 a}     \right)   \  , 
\end{eqnarray}
which apparently travels with the speed of light in spite of its massiveness. 

The origin lies in the delta function $\delta (ct - |\mbit{x}|)$ (\ref{3DkernelResult}) emerging from the discontinuity of $I$ (\ref{I Result}) between the time-, $ct > |\mbit{x}|$, and the space-like, $ct < |\mbit{x}|$, regions.  We shall call this as a light-cone singularity. (It should be emphasized that the light-cone singularity cannot become visible under the momentum representation\cite{rf:DMPF}. )

\vspace{2ex}

As a necessary consequence, any wave written as
$$
\Psi(x) = \int d^4 y {\cal D}(x,y) \psi(y)    \   ;  \qquad   {}^{\forall}\psi(y)  \ ,     
$$
must have the light-cone singularity, if ${\cal D}$ contains derivatives to some function with a discontinuity on the light-cone. The solutions of the Dirac equation meet with this, so, in the next sections, we shall study those.

\section{The Charge and the Current Density of Free Electrons}\label{Sec:The Charge Density of Free Electrons}

First let us summarize the relativistic invariant functions which participate in solving a relativistic equation. The $D$-dimensional scalar Klein-Gordon field is given as  
\begin{eqnarray}
( \dbox + \mu^2 ) \phi(x) = J(x)   \  ,   \qquad  \dbox \equiv \partial^\mu \partial_\mu  = \frac{\partial^2}{ \partial x_0^2}  - \sum_{j=1}^{D-1} \frac{\partial^2}{ \partial x_j^2}   \  ,   \label{KGeq}
\end{eqnarray}
where $\mu$ is defined by (\ref{Defmu}) and $J(x)$ is a source, complicated function of $\phi$ describing interactions. Here and hereafter the repeated indices always imply the summation. When $J=0$, the solution is
\begin{eqnarray}
\phi_0(x) = \int d^D y \Delta (x-y) \varphi(y)    \  ,      \label{KGFreeSol}
\end{eqnarray}
where $\Delta(x)$ is an invariant function defined by
\begin{eqnarray}
 & &  \hspace{9ex}  ( \dbox + \mu^2 ) \Delta(x) =  0    \  ,     \nonumber   \\
  \Delta (x) \hspace{-4ex}  & &  \equiv  - \frac{i}{ (2 \pi)^{D-1}} \int \! \! d^D k \epsilon(k_0) \delta(k^2 - \mu^2 ) {\mathrm e}^{- ikx}  \nonumber \\
  & &   =   - \frac{ic}{ (2 \pi)^{D-1}} \int \frac{ d^{D-1} \mbit{k} }{ 2 \omega_{\mbit{k}} } \left(  {\mathrm e}^{-i( \omega_{\mbit{k}} t - \mbit{k} \cdot \mbit{x} ) } - {\mathrm e}^{i( \omega_{\mbit{k}} t - \mbit{k} \cdot \mbit{x} ) }  \right)  \   ,   \  \omega_{\mbit{k} } \equiv c \sqrt{ \mbit{k}^2 + \mu^2 }   \   ,    \label{InvDeltaFn} 
\end{eqnarray}
and $\varphi(y)$ is an arbitrary function.  The notations,  
$$
k^2 \equiv  (k_0)^2 - \mbit{k}^2  \  ;  \quad    kx \equiv k_0 x_0 - \mbit{k} \cdot \mbit{x}  \ ,  \quad (x_0 \equiv ct)  \  , 
$$
with $\mbit{k}$ and $\mbit{x}$ being $(D-1)$-dimensional vector,  should be understood. When $J \neq 0$, the solution is
\begin{eqnarray}
\phi(x) = \phi_0(x) - \int d^D y  \Delta_{\rm F}(x-y) J(y)   \  ,    \label{scalarFullSolution}
\end{eqnarray}
where $ \phi_0(x)$ is (\ref{KGFreeSol}) and $\Delta_{\rm F}(x) $ is the Feynman propagator,
\begin{eqnarray}
& &  \hspace{6ex}  ( \dbox + \mu^2 ) \Delta_{\rm F}(x) =   - \delta^D (x) \left(  \equiv -\delta(x_0)\delta(x_1) \cdots \delta(x_{D-1} ) \right)    \  ,  \label{FeymanDelta}    \\
& &  \hspace{16ex}  \Delta_{\rm F}(x) \equiv \int \frac{d^D k}{ (2 \pi)^D } \frac{ {\mathrm e}^{- ikx} }{ k^2 - \mu^2 + i \epsilon}   \  .    \nonumber
\end{eqnarray}
These are shown as\cite{rf:Nakanishi}
\begin{eqnarray}
\Delta (x) \hspace{-4ex}  &  & =  - \frac{\mu^{2 \nu} }{2 (2 \pi)^\nu} \theta(x^2) \epsilon(x_0) \left( \mu \sqrt{ x^2} \right)^{-\nu} J_{-\nu} \! \left(  \mu \sqrt{ x^2}  \right)  \  ;  \hspace{10ex}    \nu \equiv \frac{D-2}{2}  \  ,   \label{D-dim.InvDelta}    \\
  \Delta_{\rm F}(x) \hspace{-4ex}  & & =    - \frac{\mu^{2 \nu}}{4 (2 \pi)^\nu } \bigg[  \theta(x^2)   \left( \mu \sqrt{ x^2} \right)^{-\nu} \left\{  J_{-\nu} \! \left(  \mu \sqrt{ x^2}  \right) - i N_{-\nu} \! \left(  \mu \sqrt{ x^2}  \right)   \right\}   \nonumber  \\
& &  \hspace{2ex} + i \left( \frac{2}{\pi}   \right) \theta(-x^2) \left( \mu \sqrt{ -x^2} \right)^{- \nu} K_{-\nu} \! \left(  \mu \sqrt{- x^2}  \right)     \bigg]  \  ;   \label{D-dim.FeynmanDelta}
\end{eqnarray}
with $x^2 \equiv (x_0)^2 - \mbit{x}^2$. By noting $J_{-n}(z) = (-)^n J_n(z), N_{-n}(z) = (-)^n N_n(z)$ and $K_{-n} (z) = K_n(z)$, they become in $D=4(\nu =1)$\cite{rf:Schweber},
\begin{eqnarray}
 & & \Delta (x) = - \frac{\epsilon(x_0) }{2 \pi} \left[  \delta (x^2) -  \frac{\mu^2}{2} \theta(x^2) \frac{ J_1 (\mu \sqrt{x^2} ) }{ \mu \sqrt{x^2} }   \right]  \  ,   \label{InvFnResult}   \\
& & \Delta_{\rm F}(x)=-\frac{1}{4\pi}\delta(x^2)+\frac{\mu^2\theta(x^2)}{8\pi}\left[\frac{J_1(\mu\sqrt{x^2})}{\mu\sqrt{x^2}}-i\frac{N_1(\mu\sqrt{x^2})}{\mu\sqrt{x^2}}\right]   \nonumber   \\
& &  \hspace{18ex}  -i\frac{\mu^2\theta(-x^2)}{4\pi^2}\frac{K_1(\mu\sqrt{-x^2})}{\mu\sqrt{-x^2}}  \  . \label{FeymanDeltaResult}
\end{eqnarray}
Note that they have a discontinuity on the light-cone. (Any relativistic invariant function does.)  In view of (\ref{KGFreeSol}) and (\ref{scalarFullSolution}), however,  there are no derivative so that we cannot have light-cone singularities for scalar fields\footnote{As for complex scalars, the current density $J_\mu(x) \equiv i e \left(  \phi^*(x) \partial_\mu \phi(x) - ( \partial_\mu \phi^*(x) ) \phi(x)  \right)$ does have a light-cone singularity.}. (We do not care the $O(1)$ delta function in (\ref{InvFnResult}) and (\ref{FeymanDeltaResult}) under the non-relativistic limit $\mu \mapsto \infty$.)

\vspace{2ex}

The free electron field $\Psi_0 (x)$ obeys the Dirac equation,
\begin{equation}
(i\dslpar -\mu)\Psi_0 (x)=0   \   ,  \qquad    \dslpar \equiv \gamma^\mu \partial_\mu  \  , \label{FreeDiracEq}
\end{equation}
where $\mu$ is (\ref{Defmu}) and $\Psi_0 (x)$ is given as
\begin{eqnarray}
\Psi_0 (x) \equiv  \left(\begin{array}{c}\varphi_0  (x)  \\
\noalign{\vspace{1ex}}
 \chi_0  (x) 
\end{array}\right)   \  ,  \label{2compoSpinor}
\end{eqnarray}
with $\varphi_0$ and $ \chi_0 $ being two component spinor.

The solution of (\ref{FreeDiracEq}) reads, 
 \begin{equation}
\Psi_0  (x)=\int d^4 y S(x-y ) \psi(y)  \   ,  \label{FreeSolByS(x)}
\end{equation}
where $S(x)$ is the invariant function for the Dirac field,
\begin{equation}
(i\dslpar -\mu)S(x)=0 \ ,   \qquad  S(x) \equiv (i \dslpar +\mu)\Delta(x) \  ,    \label{DiracDelta}
\end{equation}
with $\Delta(x) $ (\ref{InvFnResult}) and $\psi(y)$ being an arbitrary four component spinor.  From (\ref{FreeSolByS(x)}) and (\ref{DiracDelta}), $\Psi_0  (x)$ must own the light-cone singularity.

For the sake of simplicity, the initial electron configuration is assumed to be, 
\begin{equation}
\psi(y)= \delta(y_0) f( \mbit{y})  \left(\begin{array}{c} \xi_0     \\
0\end{array}\right)  \     ,   \label{InitialState}
\end{equation}
where $\xi_0$ is a constant two component spinor. Since from (\ref{InvDeltaFn}) (with $D=4$)
\begin{eqnarray}
\Delta (x) \Big|_{x_0=0} =0   \  ;   \quad  \partial_0 \Delta(x) \Big|_{x_0=0} = -\delta^3 (\mbit{x})    \  ,    \label{delta x B.C.}
\end{eqnarray}
(\ref{FreeSolByS(x)}) with (\ref{InitialState}) implies the initial condition,
\begin{eqnarray}
\Psi_0 (x_0=0 ,  \mbit{x} ) = - i  f( \mbit{y})  \left(\begin{array}{c} \xi_0     \\
0    \end{array}\right)     \  .   \label{Psi_0 InitialCond} 
\end{eqnarray}
In the following we consider three cases;
\begin{eqnarray}
f ( \mbit{y} ) \Longrightarrow \left\{ \begin{array}{l}
 \displaystyle{ f^{(3)} ( \mbit{y} ) \equiv   \frac{1}{ a^{3/4} } \exp\left(-\frac{\bm y^2}{2a}\right)   }    \  ,  \\
 \noalign{\vspace{1ex} } 
 \displaystyle{ f^{(2)} ( \mbit{y} ) \equiv  \frac{1}{ a^{1/4} } \delta(y_3)  \exp\left(-\frac{\bm y_2^2}{2a} \right)   \   ;  \qquad  { \mbit{y}_2}^2 \equiv y_1^2 + y_2^2  }  \  ,    \\
  \noalign{\vspace{1ex} } 
  \displaystyle{  f^{(1)} ( \mbit{y} ) \equiv  a^{1/4}  \delta(y_2) \delta(y_3)  \exp\left(-\frac{ y_1^2}{2a}  \right)  }  \  , 
\end{array}  \right.   \label{cases} 
\end{eqnarray}
which is called as 3-, 2-, and 1-dimensional packets respectively\footnote{\label{FN: Dimension}In terms of the length scale $L$, the dimension of a spinor field $\Psi(x)$ is $L^{-3/2}$. $\Delta(x)$ is $L^{-2}$ then $S(x)$ is $L^{-3}$ from (\ref{DiracDelta}). Hence $\psi(x)$ (\ref{FreeSolByS(x)}) is $L^{-5/2}$ so that $f(x)$(\ref{InitialState}) is $L^{-3/2}$.}. 

In view of (\ref{DiracDelta}), the solution (\ref{FreeSolByS(x)}) becomes, 
\begin{eqnarray}
   \Psi^{(k)}_0 \equiv  \! 
   \left( \!   \begin{array}{c} 
\varphi_0^{(k)}(x)     \\
\noalign{\vspace{1ex} }
\chi_0^{(k)} (x)  
\end{array}  \!   
\right) \!  = \!  \int  \! \!  d^3 \mbit{y}  \!
\left( \! \begin{array}{c} 
i\partial_0^{x}+ \mu    \\
\noalign{\vspace{1ex} }
 -i \mbit{\sigma} \cdot \Nabla^{x}   
\end{array}  \!
\right)  \! 
 \Delta(x_0 , \mbit{x}- \mbit{y}  ) f^{(k)} ( \mbit{y} )  \xi_0    \  ;  \ (k=1,2,3)   \  .      \label{Psi^(k) Definition}   
\end{eqnarray}
In the following, we shall discuss the charge and the current densities defined by\footnote{In relativity, a current 4-vector reads $J^\mu \equiv ( c \rho , \mbit{J} )$ in the MKS unit. The Dirac particle $\Psi$ induces a current $J^\mu \equiv e \overline{\Psi} \gamma^\mu \Psi$ with $\overline{\Psi} \equiv \Psi^\dagger \gamma^0 $.}
\begin{eqnarray}
\rho^{(k)} \hspace{-4ex} & & \equiv \frac{e}{c} {\Psi_0^{(k)} }^{\dagger} \Psi^{(k)} _0  =  \frac{e}{c} \! \left(    \!  {\varphi^{(k)}_0}^\dagger  \!  (x)   \varphi^{(k)}_0  \!  (x)  \!  + {\chi^{(k)}_0 }^\dagger  \!  (x) \chi^{(k)}_0  \!  (x)  \!     \right)  ,  \nonumber   \\
  \mbit{J}^{(k)}  \hspace{-4ex} & &  \equiv e {\Psi_0^{(k)} }^{\dagger} \! \gamma^0 \mbit{\gamma} \ \Psi^{(k)}_0 \!   \!  =e \! \left(  \!  {\varphi^{(k)}_0}^\dagger  \!  (x)  \mbit{\sigma} \chi^{(k)}_0  \!  (x)  \!  + {\chi^{(k)}_0 }^\dagger  \!  (x)  \mbit{\sigma} \varphi^{(k)}_0  \!  (x)    \! \right)   ,   \label{ChargeCurrentDef}
\end{eqnarray}

\vspace{1ex}

Now take the non-relativistic limit $c \mapsto \infty \ (\mu \mapsto \infty)$ to find that $\Delta(x)$ (\ref{InvFnResult}) reduces to
\begin{eqnarray}
\Delta(x) \stackrel{\mu \mapsto \infty}{\approx}   \frac{\mu^2 \epsilon(x_0) }{4 \pi} \theta(x^2) \frac{ J_1 (\mu \sqrt{x^2} ) }{ \mu \sqrt{x^2} }     \  ,  \label{mu infty Delta}
\end{eqnarray}
whose derivatives are calculated as follows: first note 
\begin{eqnarray}
\partial_0 \left( \epsilon(x_0) \theta(x^2) \right) =  2 |x_0|\delta(x^2)   \  ,   \label{partial_0 epsilon(x_0)}
\end{eqnarray}
by use of 
$$
\partial_0 \epsilon (x_0) = 2 \delta (x_0) \ ; \quad  \delta(x_0) \theta(x^2)   =0  \ ;  \quad  x_0 \epsilon(x_0)= |x_0|  \  ,
$$
as well as $\partial_0 = 2x_0 \partial/\partial x^2$ and 
\begin{eqnarray}
& &  \hspace{2ex}    \frac{\partial}{\partial x^2} \theta(\pm x^2) = \pm   \delta (x^2)  \   .     \label{theta-delta}   
\end{eqnarray}
Thus
\begin{eqnarray*}
\partial_0 \Delta(x)=  \frac{\mu^2  }{4 \pi}2 |x_0|\delta(x^2)  \frac{ J_1 (\mu \sqrt{x^2} ) }{ \mu \sqrt{x^2} }   +   \frac{\mu^2  }{4 \pi}\epsilon(x_0) \theta(x^2) 2x_0  \frac{\partial}{\partial x^2} \left( \frac{ J_1 (\mu \sqrt{x^2} ) }{ \mu \sqrt{x^2} } \right)   \  , 
\end{eqnarray*}
whose second term reads
\begin{eqnarray*}
 \mu^2 \frac{\partial}{\partial x^2} \left( \frac{ J_1 (\mu \sqrt{x^2} ) }{ \mu \sqrt{x^2} } \right) \stackrel{z \equiv  \mu \sqrt{x^2} }{=} \mu^2 \frac{\partial z}{\partial x^2} \frac{d}{dz} \left( \frac{ J_1 (z ) }{ z} \right)  \stackrel{(\ref{BesselDifferntiation})}=  - \frac{ \mu^2}{2 x^2} J_2 (\mu \sqrt{x^2} ) = O ( \mu^{3/2} )   \   ,
\end{eqnarray*}
where $ J_n (z)  \stackrel{ z \mapsto \infty}{=} O (1/ \sqrt{z})$ (\ref{asymExp Bessel}) has been considered in the final expression. Meanwhile the first term becomes, with the aid of $J_1 (z )/ z\big|_{z=0} = 1/2$ (\ref{J_1/z I_1/z at z=0}), yielding to
\begin{eqnarray}
\partial_0 \Delta(x) = \frac{ \mu^2}{4 \pi} |x_0| \delta (x^2) + O \! \left(  \mu^{3/2} \right) \  .   \label{partial_x_0 Delta(x)}
\end{eqnarray}
Similarly by $\partial_k = - 2x_k \partial /\partial x^2$ and (\ref{theta-delta})
\begin{eqnarray}
\partial_k  \Delta(x) =  - \frac{ \mu^2}{4 \pi} x_k \epsilon(x_0) \delta (x^2) + O \! \left(  \mu^{3/2} \right) \  .   \label{partial_x_k Delta(x)}
\end{eqnarray}
From these, we can convince that the leading terms in the non-relativistic approximation are nothing but the light-cone singularities. 
Therefore by noting that
\begin{eqnarray}
\delta(x^2) = \frac{1}{2 | \mbit{x} | } \Big[    \delta (x_0 -| \mbit{x} |) + \delta (x_0 + | \mbit{x} |)  \Big]   \  ,  \label{deltaFn Fomula}
\end{eqnarray}
as well as $\mu  \Delta(x)  = O \! \left(  \mu^{3/2} \right) $,  (\ref{Psi^(k) Definition}) reads as
\begin{eqnarray}
 \Psi^{(k)}_0 =  \frac{i\mu^2  }{8\pi }  \! \! \! \int \! d^3\mbit{y} \delta(x_0 -|\bm y|)  \left(
\begin{array}{c} 
1   \\
\noalign{\vspace{1ex} }
\displaystyle{  \frac{ - \mbit{y} \cdot \mbit{\sigma} }{ |  \mbit{y} | } }
\end{array}
\right)   f^{(k)} (\mbit{y} + \mbit{x} ) \xi_0 + O \! \left(  \mu^{3/2} \right)     \   ,  \label{FreeBeforeFinal} 
\end{eqnarray}
where we have noticed $x_0 >0$ and made a shift $ \mbit{y} \mapsto \mbit{y} + \mbit{x}$ .

Now proceed to individual cases: from (\ref{cases}) and (\ref{FreeBeforeFinal}) the 3-dimensional solution is
\begin{eqnarray}
\Psi_0^{(3)} = \frac{i\mu^2 }{8\pi a^{3/4} } \exp \left( - \frac{ \mbit{x}^2}{2a }   \right)  
\left(
\begin{array}{c} 
1   \\
\noalign{\vspace{1ex} }
\displaystyle{  \frac{a}{x_0}  \mbit{\sigma} \cdot \Nabla  }
\end{array}
\right)    I^{(3)}(x_0, \mbit{x} ) \xi_0   \  ,
\end{eqnarray}
where
\begin{eqnarray}
I^{(3)} (x_0, \mbit{x} ) \equiv   \int d^3 \mbit{y} \delta(x_0 -|\bm y|) \exp \left(  - \frac{ \mbit{x} \cdot \mbit{y} }{a} -  \frac{ \mbit{y}^2 }{2 a}  \right)  \ ,   \hspace{8ex}    \label{3DIntegral}     
\end{eqnarray}
which becomes with the aid of the polar coordinates
\begin{eqnarray}
I^{(3)} (x_0, \mbit{x} ) = \frac{2 \pi a x_0}{| \mbit{x} | }\exp \left( -  \frac{ x_0^2 }{2 a}  \right)
\left[  \exp \left(  \frac{| \mbit{x} | x_0 }{ a}  \right)  - \exp \left( - \frac{| \mbit{x} | x_0 }{ a}  \right)  \right]  \   .   \label{3DIntegral Result}  
\end{eqnarray}
Hence
\begin{eqnarray}
\Psi_0^{(3)} = \frac{i\mu^2 a^{1/4} x_0 }{4 | \mbit{x} | } \Bigg[  
 \left(
\begin{array}{c} 
1   \\
\noalign{\vspace{1ex} }
\displaystyle{   1- \frac{a}{x_0 |\mbit{x}| }   }
\end{array}
\right)  
\exp \left( - \frac{ (x_0 - |\mbit{x}| )^2}{2a }   \right)  \hspace{16ex}   \nonumber  \\
\mp 
\left(
\begin{array}{c} 
1   \\
\noalign{\vspace{1ex} }
\displaystyle{   1+ \frac{a}{x_0 |\mbit{x}| }   }
\end{array}
\right) \exp \left( - \frac{ (x_0 + |\mbit{x}| )^2}{2a }   \right)  \Bigg]  \left(
\begin{array}{c} 
 \xi_0     \\
\noalign{\vspace{1ex} }
 \mbit{n} \cdot \mbit{\sigma}     \xi_0
\end{array}
\right)  \  ,  \quad   \mbit{n} \equiv \frac{\mbit x}{| \mbit x |}    \  .  \label{3Dfree phi chi}  
\end{eqnarray}
When $x_0,  |\mbit{x} |  \gg  \sqrt{a} $, we have
\begin{eqnarray*}
\Psi_0^{(3)}  \approx  \frac{i\mu^2  a^{1/4} x_0 }{4  |\mbit{x} | }\exp\left(-\frac{(x_0 -| \mbit{x} |)^2}{2a}\right) \left(
\begin{array}{c}
  \xi_0   \\
  \mbit{n} \cdot \mbit{ \sigma} \xi_0 
 \end{array}
\right)    \  , 
\end{eqnarray*}
which further turns out to be
\begin{eqnarray}
\Psi_0^{(3)}   \approx  \frac{i\mu^2  a^{1/4}  }{4  }\exp\left(-\frac{(x_0 -| \mbit{x} |)^2}{2a}\right) \left(
\begin{array}{c}
  \xi_0   \\
  \mbit{n} \cdot \mbit{ \sigma} \xi_0 
 \end{array}
\right)    \  ,   \label{3Dfree Psi}
\end{eqnarray}
around the peak,
\begin{eqnarray}
x_0- |\mbit{x} | \approx 0 \  ,  \label{mostProbabArea}
\end{eqnarray}
which apparently travels with the speed of light.

Insert (\ref{3Dfree Psi}) into (\ref{ChargeCurrentDef}) to find
\begin{eqnarray} \rho^{(3)}  =  \frac{2e}{c} \left(\frac{\mu^2 a^{1/4}   }{4 }\right)^{ \! \! 2}  \exp \! \left(-\frac{(x_0-| \mbit{x} |)^2}{a}\right)  \xi_0^\dagger \xi_0   \ ,   \quad     \mbit{J}^{(3)}  =  c \mbit{n} \rho^{(3)}     \ ,    \label{3DFreeCurDensity}
\end{eqnarray}
where use has been made of the anti-commutation relations $ \left\{ \sigma_j , \sigma_k \right\} = 2 \delta_{jk} $ in $\mbit{J}^{(3)}$.  Those travel with the speed of light.

It would be easier to prepare packets restricted in the 2- or 1-dimensional region.  The former reads, by inserting (\ref{cases}) into (\ref{FreeBeforeFinal}), as
\begin{eqnarray}
\Psi_0^{(2)}  = \frac{i\mu^2 }{8\pi a^{1/4} } \exp \left( - \frac{{ \mbit{x}_2}^2}{2a }   \right) \left(
\begin{array}{c} 
1   \\
\noalign{\vspace{1ex} }
\displaystyle{  \frac{a}{x_0}  \mbit{\sigma}_2 \cdot \frac{\partial }{\partial \mbit{x}_2 }  }
\end{array}
\right)  I^{(2)} (x_0, \mbit{x}_2 )  \xi_0    \  .   \label{2Dfree phi chi} 
\end{eqnarray}
where $\mbit{\sigma}_2 \equiv ( \sigma_1, \sigma_2), \mbit{x}_2 \equiv (x_1, x_2) $ and
\begin{eqnarray}
I^{(2)} (x_0, \mbit{x}_2 ) \equiv \int d^2 \mbit{y}_2 \delta(x_0 -| \mbit{y}_2 | ) \exp \left(  - \frac{ \mbit{x}_2 \cdot \mbit{y}_2}{a} -  \frac{ {\mbit{y}_2}^2 }{2 a}  \right)   \   .   \label{2DIntegral}
\end{eqnarray}
Here we have put $x_3 =0$, since the observation should also be made in the $x_3=0$ plane. (\ref{2DIntegral}) becomes under the polar coordinates, to
\begin{eqnarray*}
I^{(2)} (x_0, \mbit{x}_2 )\hspace{-4ex} & &  =  \int_0^\infty dy y \int_0^{2 \pi} d \theta  \delta ( x_0 -y ) \exp \left( - \frac{| \mbit{x}_2 |y }{a} \cos \theta  -  \frac{ y^2 }{2 a}     \right)    \\
 & & = x_0 \exp\left(  - \frac{x_0^2}{2 a} \right)  \int_0^{2 \pi} d \theta  \exp  \left(  - \frac{| \mbit{x}_2 |x_0 }{a} \cos \theta      \right) \  . 
\end{eqnarray*}
When $| \mbit{x}_2 |$, $ x_0 \gg \sqrt{a}$, the saddle point $\theta= \pi$ gives a asymptotic value such that
\begin{eqnarray}
I^{(2)} (x_0, \mbit{x}_2 ) \approx   \sqrt{  \frac{ 2 \pi a x_0 }{ | \mbit{x}_2 |}  } \exp \left( - \frac{  x_0^2  }{ 2a }  + \frac{| \mbit{x}_2 | x_0 }{a}   \right)  \  .  
\end{eqnarray}
Thus (\ref{2Dfree phi chi}) reads
\begin{eqnarray}
\Psi_0^{(2)}   \approx  \frac{ i\mu^2 a^{1/4} }{4 \sqrt{ 2\pi }}    \exp \left( - \frac{ ( x_0 - | \mbit{x}_2 |)^2}{2a }   \right) \left(
\begin{array}{c}
\xi_0  \\
 \noalign{\vspace{1ex} }
\mbit{n}_2 \cdot   \mbit{\sigma}_2 \xi_0   
\end{array}
\right)    \  .    \label{2DfreeResult phi chi}   
\end{eqnarray}
Here the final expressions has been obtained by putting $x_0 / | \mbit{x}_2 | \mapsto 1$ in the coefficient; since the peak is now given as
\begin{equation}
x_0 - | \mbit{x}_2 | \approx 0  \  ,    \label{2DmostProbArea}
\end{equation}
whose velocity is again the light speed.  The charge and the current densities (\ref{ChargeCurrentDef}) read 
\begin{eqnarray}
 \rho^{(2)}  = \frac{2e}{c}\left( \frac{ \mu^2 a^{1/4} }{4 \sqrt{ 2\pi } }  \right)^{ \! \! 2}  \exp \! \left(-\frac{(x_0-| \mbit{x}_2 |)^2}{a}\right)  \xi_0^\dagger \xi_0    \ ,    \quad   \mbit{J}^{(2)}  = \left(   c \mbit{n}_2 \rho^{(2)} ,  0  \right)       \   .      \label{2DFreeCurrnDensity}
\end{eqnarray}

\vspace{2ex}

Finally we discuss the 1-dimensional case written, after putting $x_2=0 , x_3=0$, as
\begin{eqnarray}
 \Psi_0^{(1)} = \frac{i\mu^2 a^{1/4} }{8\pi } \exp \left( - \frac{ x_1^2}{2a }   \right)  \left( \! 
\begin{array}{c}
1 \\
 \noalign{\vspace{1ex} }
\displaystyle{  \frac{ a }{x_0}  \sigma_1 \partial_1  } 
\end{array} \! 
\right)  I^{(1)}( x_0, x_1 )  \xi_0   \  ,  \label{1Dfree phi chi}
\end{eqnarray}
where
\begin{eqnarray}
I^{(1)} (x_0, x_1 )  \equiv \int_{-\infty}^\infty d y_1  \delta(x_0 -| y_1| ) \exp \left(  - \frac{ x_1 y_1}{a} -  \frac{ y_1^2 }{2 a}  \right)  \ ,   
\end{eqnarray}
which turns out, by recalling $x_0 > 0$, to be
\begin{eqnarray}
I^{(1)} (x_0, x_1 ) \hspace{-4ex} & & = \left[  \int_{-\infty}^0 d y_1  \delta(x_0 + y_1 )  + \int_{0}^\infty d y_1  \delta(x_0 - y_1 )  \right] \exp \left(  - \frac{ x_1 y_1}{a} -  \frac{ y_1^2 }{2 a}  \right)  \nonumber \\
\hspace{-4ex} & &  = \bigg[ \exp \left(  \frac{ x_0 x_1 }{a} \right)  +  \exp \left(  - \frac{ x_0 x_1 }{a} \right)  \bigg]  \exp \left(    -  \frac{ x_0^2 }{2 a}  \right)   \  .  
\end{eqnarray}
Thus 
\begin{eqnarray}
   \Psi_0^{(1)}   = \frac{i\mu^2 a^{1/4} }{8\pi } \!  \left[  \exp \!  \left(  \! - \frac{(x_0 - x_1)^2 }{2a }  \right) \!  \pm  \exp \! \left( \!  - \frac{(x_0 + x_1)^2 }{2a }   \right) \!  \right] \! \!  
 \left( \!  \!  \! 
\begin{array}{c}
\xi_0  \\
 \noalign{\vspace{1ex} }
 \sigma_1 \xi_0  
\end{array} \!  \! 
\right)   \  .   \label{1DFree phi chiResult2}
\end{eqnarray}
When $x_0, x_1 \gg \sqrt{a}$, it becomes
\begin{eqnarray}
 \Psi_0^{(1)}  \approx  \frac{i\mu^2 a^{1/4} }{8\pi } \!  \exp \!  \left(  \! - \frac{(x_0 - x_1)^2 }{2a }  \right) \left( \!  \!  \! 
\begin{array}{c}
\xi_0  \\
 \noalign{\vspace{1ex} }
 \sigma_1 \xi_0  
\end{array} \!  \! 
\right)  \  ,  \label{1DFree phi chiResult Final}
\end{eqnarray}
whose peak is
\begin{equation}
x_0 - x_1 \approx 0  \  ,   \label{1DmostProbArea}
\end{equation}
around which the charge and the current densities (\ref{ChargeCurrentDef}) read
\begin{eqnarray}
\rho^{(1)}   =\frac{2e}{c}\left(  \frac{ \mu^2 a^{1/4} }{8 \pi }  \right)^{\! \! 2} \exp \! \left(-\frac{(x_0-x_1)^2}{a} \right)  \xi_0^\dagger \xi_0    \ ,   \quad    \mbit{J}^{(1)} = \left(  c \rho^{(1)}  ,  0  , 0  \right)      \  .    \label{1DFreeCurDensity}   
\end{eqnarray}
In view of (\ref{3DFreeCurDensity}), (\ref{2DFreeCurrnDensity}) and (\ref{1DFreeCurDensity}), the free electrons travel with the speed of light in the non-relativistic world\footnote{Note that the relation between the current and the charge density $\mbit{J} = c \mbit{n} \rho$ in (\ref{3DFreeCurDensity}), (\ref{2DFreeCurrnDensity}), and (\ref{1DFreeCurDensity}) is merely a kinematical consequence by taking the upper component $\xi_0$ only. \label{footnote:J=c rho}}.

\section{Electrons in a Laboratory}\label{Sec:The Charge and the Current Density for Electrons in a Laboratory}

In a realistic situation, electrons interact with the electromagnetic field $A_\mu(x)$ such that
\begin{equation}
(i\dslpar-\mu)\Psi(x)= \frac{ e }{ \hbar c} \gamma^{\mu}A_{\mu}(x) \Psi(x) \  ,   
\end{equation}
whose solution is
\begin{eqnarray}
\Psi(x) = \Psi_0  (x) +  \int d^4 y S_{\rm F} (x -y) e\gamma^{\mu}A_{\mu}(y) \Psi(y)   \ ,   \label{generalSolu0}
\end{eqnarray}
where $\Psi_0  (x)$ is the free field discussed in the previous section and $S_{\rm F} (x)$ is the Feynman propagator for the Dirac field, 
\begin{equation} 
  (i\dslpar-\mu) S_{\rm F}  (x) =  \delta^4 (x)   \   ,  \quad  S_{\rm F} (x) \equiv (i\dslpar+\mu) \Delta_{\rm F} (x) \  ,  
\end{equation}
with $ \Delta_{\rm F}(x) $ given in (\ref{FeymanDeltaResult}).

The interaction is assumed to take place for a finite interval, giving
\begin{eqnarray}
\frac{ e }{ \hbar c} \gamma^{\mu}A_{\mu}(y) \Psi(y) \equiv  h^{(k)}(y) \left(\begin{array}{c} \xi_1  \\
0\end{array}\right) \  ,   \quad  ( k=1,2, 3)  \  ,     \label{H Definition}
\end{eqnarray}
where 3-, 2-, and 1-dimensional packets are
\begin{eqnarray}
\begin{array}{l}
 \displaystyle{ h^{(3)}(y_0, \mbit{y}) \equiv  \frac{1}{(\tau + b)^{1/4} \tau^{1/4}  b^{3/4} } \exp\left( - \frac{y_0^2 }{2 \tau } -\frac{\bm y^2}{2b}\right) }    \  ,  \\
 \noalign{\vspace{1ex} } 
 \displaystyle{ h^{(2)}(y_0, \mbit{y}) \equiv   \frac{1}{ (\tau + b)^{1/4}  \tau^{1/4}  b^{1/4} } \delta(y_3)  \exp\left( - \frac{y_0^2 }{2 \tau } 
 -  \frac{ {\bm y}_2^2}{2b}\right)  }   ;  \quad  \mbit{y}_2^2 \equiv y_1^2 + y_2^2    \  ,    \\
 \noalign{\vspace{1ex} } 
  \displaystyle{  h^{(1)}(y_0 , \mbit{y})  \equiv    \frac{b^{1/4} }{(\tau + b)^{1/4} \tau^{1/4}  }  \delta(y_2) \delta(y_3)  \exp\left( - \frac{y_0^2 }{2 \tau } -\frac{y_1^2}{2b}\right)  }    \  ,
\end{array}                   \label{sourceDef} 
\end{eqnarray}
with $ \xi_1$ being a constant two component spinor.  (Note that the dimension of $h(y)$ is $L^{-5/2}$. See the footnote \ref{FN: Dimension}.) If we write
\begin{equation}
\Psi_h^{(k)}  \!  \equiv  \! 
\left( \! 
\begin{array}{c} \varphi_h^{(k)}  (x)  \\
\noalign{\vspace{1ex}} 
\chi_h^{(k)}  (x) 
 \end{array}  \!  
 \right)  \! 
  \equiv \int d^4y \left( \! 
\begin{array}{c} i \partial_0^{x}+ \mu   \\
\noalign{\vspace{1ex}} 
 -i \mbit{\sigma} \cdot \Nabla^{x}    
 \end{array}  \!  
 \right)  \!   \Delta_{\rm F} (x-y ) h^{(k)}(y )  \xi_1   \  ;  \quad    (k=1,2,3)   \  ,       \label{Psi_HDef}
\end{equation}
the general solution (\ref{generalSolu0}) reads as
\begin{equation}
\Psi^{(k)} =\Psi_0^{(k)} +\Psi_h^{(k)}  \   ,      \label{generalSolu1} 
\end{equation}
with $\Psi_0^{(k)}$ given by (\ref{Psi^(k) Definition}).

In the non-relativistic limit $\mu \mapsto \infty$, $\Delta_{\rm F}(x)$ (\ref{FeymanDeltaResult}) reduces to
\begin{eqnarray}
 \Delta_{\rm F}(x) \hspace{-4ex}  & & \stackrel{\mu \mapsto \infty}{\approx}    \frac{\mu^2\theta(x^2)}{8\pi}   \left[   \frac{J_1(\mu\sqrt{x^2})}{\mu\sqrt{x^2}} -  i \frac{N_1(\mu\sqrt{x^2})}{\mu\sqrt{x^2}} \right]  -i\frac{\mu^2\theta(-x^2)}{4\pi^2}\frac{K_1(\mu\sqrt{-x^2})}{\mu\sqrt{-x^2}}   \nonumber  \\
& &  \hspace{2ex} =  \frac{\mu^2\theta(x^2)}{8\pi}\frac{ H_1^{(2)} (\mu\sqrt{x^2})}{\mu\sqrt{x^2}}   -i\frac{\mu^2\theta(-x^2)}{4\pi^2}\frac{K_1(\mu\sqrt{-x^2})}{\mu\sqrt{-x^2}}  \  ,   
\end{eqnarray}
where we have utilized (\ref{H^(2) Definition}) to the final expression whose form reminds us (\ref{I Result}) in sec.\ref{Sec:ToyModel}.

The derivative reads
\begin{eqnarray*}
  \quad  \partial_\mu \Delta_{\rm F}(x) = 2x_\mu  \frac{\mu^2}{8\pi} \delta (x^2) \left[   \frac{J_1(\mu\sqrt{x^2})}{\mu\sqrt{x^2}} -\frac{ 2 i}{\pi}   \!  \left( \! \frac{\pi}{2} \frac{N_1(\mu\sqrt{x^2})}{\mu\sqrt{x^2}}   -  \frac{K_1(\mu\sqrt{-x^2})}{\mu\sqrt{-x^2}} \!  \right) \!  \right] \! \! + O \! \left( \mu^{3/2} \right)  \  ,  
\end{eqnarray*}
where use has been made of  (\ref{theta-delta}) and the asymptotic behavior (\ref{asymExp Bessel}). Noting (\ref{J_1/z I_1/z at z=0}) and (\ref{N_1/z -K_1/z}) in Appendix \ref{AppendixA}
we have the light-cone singularity, 
\begin{eqnarray}
 \partial_\mu \Delta_{\rm F}(x) = x_\mu  \frac{i\mu^2}{4\pi^2} \delta (x^2) + O \! \left( \mu^{3/2} \right)  \  .  \label{deltaF singularity} 
\end{eqnarray}

Contrary to the previous situation, we now need both terms in (\ref{deltaFn Fomula}), since $y_0$ cannot always be positive, so that
\begin{eqnarray}
   (i\partial_0^x+\mu)\Delta_{\rm F}(x -y)   
=\frac{-\mu^2}{8 \pi^2} \!  \Big[  \delta(x_0 - y_0 -|\mbit{x} - \mbit{y}|) \! - \!  \delta(x_0 - y_0 +|\mbit{x} - \mbit{y}|) \Big] \!    + O \! \left( \mu^{3/2} \right)   ,   \label{deltaFn Fomula2}    \\
-i\mbit{\sigma} \! \cdot \! \Nabla^x \! \! \Delta_{\rm F}  (  x \! - \! y)  \!  \! 
 = \! \frac{-\mu^2 \mbit{\sigma} \! \cdot  \! ( \mbit{x} \! - \! \mbit{y}) }{8\pi^2 | \mbit{x} - \mbit{y} |}    \Big[   \delta( x_0 \! - \! y_0 \!  - \! |\mbit{x} - \mbit{y}|) \! + \! \delta(x_0 \! - \!  y_0 \! + \! |\mbit{x} - \mbit{y}|) \! \Big] \! \! + \! O \! \left( \! \mu^{3/2} \! \right) 
    .   \label{deltaFn Fomula3} 
\end{eqnarray}
Then (\ref{Psi_HDef}) turns out to be 
\begin{eqnarray}
\Psi_h^{(k)} = \frac{-\mu^2}{8 \pi^2}\! \! \int \! \! d^3 \mbit{y}   \left(     h^{(k)}(x_0 - |\mbit{x} - \mbit{y}|, \mbit{y} ) \mp    h^{(k)}(x_0 + |\mbit{x} - \mbit{y}|, \mbit{y} )   \right) 
  \left(
\begin{array}{c} 
1   \\
\noalign{\vspace{1ex} }
\displaystyle{  \frac{ - \mbit{y} \cdot \mbit{\sigma} }{ |  \mbit{y} | } }
\end{array}
\right) \xi_1  \  ,  \label{PsiWithSource} 
\end{eqnarray}
where a shift $\mbit{y} \mapsto \mbit{y}+ \mbit{x}$ has been made.

Let us exam individual cases: with the 3-dimensional packet (\ref{PsiWithSource}) read, 
\begin{eqnarray}
\Psi^{(3)}_h = \frac{-\mu^2}{8 \pi^2} \left(  \frac{1 }{\tau b^3 (\tau +b) } \right)^{1/4} \! \! \!  \! \!  \exp \! \left( - \frac{x_0^2}{2 \tau} - \frac{\mbit{x}^2}{2 b}  \right) \! \!   \left(  \! \! 
\begin{array}{c} 
\displaystyle{ \tau \partial_0       }  \\
\noalign{\vspace{1ex} }
\displaystyle{ b  \mbit{\sigma} \cdot \Nabla  }
\end{array}  \! \! 
\right) I_h^{(3)} (x_0, \mbit{x} )    \xi_1  \ ,    \label{3D withH phi chi Definition}
\end{eqnarray}
where
\begin{eqnarray}
I_h^{(3)} (x_0, \mbit{x} ) \hspace{-4ex} & &  \equiv  \! \int \! \!  \frac{ d^3 \mbit{y} }{ | \mbit{y} | } \left\{  \exp \! \left(  \frac{x_0 | \mbit{y} | }{ \tau}  \right) + \exp \! \left( - \frac{x_0 | \mbit{y} | }{ \tau}  \right)  \right\}  \exp \! \left( - \frac{(\tau + b) \mbit{y}^2}{2 \tau b} - \frac{\mbit{x}\cdot \mbit{y} }{ b}  \right)  \nonumber  \\
& & = 2  \int \! \!  \frac{ d^3 \mbit{y} }{ | \mbit{y} | } \exp \! \left( - \frac{(\tau + b) \mbit{y}^2}{2 \tau b}  \right) \cosh \left(  \frac{x_0  }{ \tau}  | \mbit{y} | \right)   \exp \! \left(  - \frac{\mbit{x}\cdot \mbit{y} }{ b}  \right) \   ,    \label{I_N^(3)}
\end{eqnarray}
which, with the aid of the polar coordinates and the error function\cite{rf:Ryz6},
\begin{eqnarray}
\mathrm{erf}(x)\equiv\frac{2}{\sqrt\pi}\int_0^{x}dt\exp(-t^2) \  ,    \label{errorFunc Def}
\end{eqnarray}
yields to
\begin{eqnarray}
  I_h^{(3)} (x_0, \mbit{x} ) =  \sqrt{ \frac{( 2 \pi b)^3 \tau }{\tau + b}  } \frac{1}{| \mbit{x} | } \left[ {\rm erf} \!  \left(  \frac{ b x_0 + \tau| \mbit{x} |}{ \sqrt{2 \tau b (\tau + b) } } \right) \exp \! \left(  \frac{( b x_0 + \tau| \mbit{x} | )^2}{2 \tau b (\tau + b)}   \right) \right.   \nonumber \\
  \left. -  {\rm erf} \!  \left(  \frac{ b x_0 - \tau| \mbit{x} |}{ \sqrt{2 \tau b (\tau + b) } } \right) \exp \! \left(  \frac{( b x_0 - \tau| \mbit{x} | )^2}{2 \tau b (\tau + b)}   \right)    \right]     \  .   \label{I_N^(3) Result}
\end{eqnarray}
Applying the differentiations, in (\ref{3D withH phi chi Definition}) and assuming $x_0, |\mbit{x}| \gg \sqrt{\tau} , \sqrt{b}$, we obtain
\begin{eqnarray}
\Psi_h^{(3)} \approx \frac{-\mu^2}{ 2 \sqrt{2 \pi} } \left(  \frac{\tau b^3 }{(\tau +b )^3} \right)^{\! \! 1/4}  \exp \! \left(  - \frac{(  x_0 - | \mbit{x} | )^2}{2 (\tau + b)}   \right) \left(
\begin{array}{c}
  \xi_1   \\
  \mbit{n} \cdot \mbit{ \sigma} \xi_1 
 \end{array}
\right)   \  ,   \label{3Dint Psi}
\end{eqnarray}
around the peak (\ref{mostProbabArea}). (Details are relegated to Appendix \ref{AppendixB}.)

Therefore in view of $\Psi^{(3)}_0$(\ref{3Dfree Psi}) and $\Psi^{(3)}_h$(\ref{3Dint Psi}),  the total $\Psi$(\ref{generalSolu1}) is obtained as
\begin{eqnarray}
\Psi^{(3)} =  \frac{ i\mu^2 a^{1/4} }{4}  \exp \! \left(  - \frac{\tau + a + b }{4a (\tau + b)} (  x_0 - | \mbit{x} | )^2   \right) 
\left(
\begin{array}{c}
  \xi    \\
  \mbit{n} \cdot \mbit{ \sigma} \xi  
 \end{array}
\right)   \   ,  
\end{eqnarray}
where we have introduced a two component spinor
\begin{eqnarray}
  \xi  \hspace{ -4ex} & & \equiv   \exp \! \left(  - \frac{\tau - a + b }{4a (\tau + b)}  (  x_0 - | \mbit{x} | )^2    \right)  \xi_0    \nonumber  \\
& &  \hspace{2ex} 
 + i \sqrt{  \frac{2} { \pi} }\left(  \frac{\tau b^3 }{a (\tau +b )^3}  \right)^{\! \! 1/4}   \exp \! \left(  \frac{ \tau - a + b  }{4a (\tau + b)}  (  x_0 - | \mbit{x} | )^2  \right)  \xi_1   \   . 
\end{eqnarray}
The charge and the current densities are 
\begin{eqnarray}
\rho_{\rm tot}^{(3)}  \equiv \frac{e}{c} {\Psi^{(3)} }^{\dagger}\Psi^{(3)}  = \frac{2e}{c} \left(  \frac{\mu^2 a^{1/4} }{4}   \right)^2 \exp \! \left(  - \frac{\tau + a + b }{2a (\tau + b)} (  x_0 - | \mbit{x} | )^2   \right)  \xi^\dagger \xi  \hspace{12ex}   \nonumber \\
=  \frac{2e}{c} \!  \! \left(  \! \frac{\mu^2 a^{1/4} }{4}  \right)^{ \! \! \! 2} \! \left[   \exp\left( \! -\frac{(x_0-|\mbit x|)^2}{a}\right)\xi_0^{\dagger}\xi_0  +   \frac{2}{\pi}  \sqrt{  \frac{\tau b^3 }{a(\tau +b )^3} } 
\exp \left( \! -\frac{(x_0-|\mbit x|)^2}{2(\tau+b)}\right)  \xi_1^{\dagger} \xi_1    \right.  \hspace{-2ex}  \nonumber \\
 +\left.  i \sqrt{\frac{2}{\pi}} \left(  \frac{\tau b^3 }{a (\tau +b )^3}  \right)^{\! \! 1/4} \hspace{-2ex}   \exp \! \left(  - \frac{ \tau + a + b  }{2 a (\tau + b)}  (  x_0 - | \mbit{x} | )^2  \right)  (\xi_0^{\dagger}\xi_1- \xi_1^{\dagger}\xi_0)\right]       \   ,    \nonumber  \\
\mbit{ J}_{\rm tot}^{(3)}  \equiv  e {\Psi^{(3)} }^{\dagger} \gamma^0\mbit\gamma \Psi^{(3)}  = c \mbit n  \rho_{\rm tot}^{(3)}    \   .   \hspace{20ex}   \label{3DFullChargeDensity}
\end{eqnarray}
Since each term has a peak at the light-cone, the electron signal traveling with the light speed would be observed.

Next we consider the 2-dimensional case: again we restrict $\mbit x$ to be $x_3=0$ so that (\ref{PsiWithSource}) with (\ref{sourceDef}) are found as
\begin{eqnarray}
\Psi_h^{(2)} = \frac{-\mu^2}{8 \pi^2} \left(  \frac{1 }{ \tau b (\tau +b )  } \right)^{1/4}   \exp \! \left( - \frac{ x_0^2  }{2 \tau} - \frac{\mbit{x}_2^2}{2 b}  \right)  \left(
\begin{array}{c} 
\displaystyle{ \tau \partial_0     }  \\
\noalign{\vspace{1ex} }
\displaystyle{ b \mbit{\sigma}_2 \cdot \frac{\partial }{\partial \mbit{x}_2 }  }
\end{array}
\right) I_h^{(2)} (x_0, \mbit{x}_2 )    \xi_1  \  ,     \label{2D phi chi}  
\end{eqnarray}
where
\begin{eqnarray}
I_h^{(2)} (x_0, \mbit{x}_2 ) \equiv 2 \int \frac{d^2 \mbit{y}_2}{ |\mbit{y} |}  \exp \left(  - \frac{(\tau +b )}{2 \tau b} \mbit{y}_2^2 - \frac{\mbit{x}_2 \cdot \mbit{y}_2 }{b}  \right) \cosh \left(  \frac{x_0 |\mbit{y} |}{ \tau}  \right)   \  ,   \label{2DInt withSouce}
\end{eqnarray}
which yields
\begin{eqnarray}
I_h^{(2)} (x_0, \mbit{x}_2 )\approx \sqrt{\frac{( 2 \pi )^2 \tau b^2}{ | \mbit{x}_2 | ( \tau | \mbit{x}_2 | +  b x_0 ) }}  \exp \left(  \frac{( \tau | \mbit{x}_2 |  + b x_0 )^2}{2 \tau b (\tau + b) }   \right) \hspace{10ex}  \nonumber  \\
+ \sqrt{\frac{( 2 \pi )^2 \tau b^2}{ | \mbit{x}_2 | ( \tau | \mbit{x}_2 | -  b x_0 ) }}  \exp \left(  \frac{( \tau | \mbit{x}_2 |  -  b x_0 )^2}{2 \tau b (\tau + b) }   \right)    \  .   \label{I^(2)_H Result}
\end{eqnarray}
Therefore when $x_0, |\mbit{x}_2| \gg \sqrt{\tau}, \sqrt{b}$
\begin{eqnarray}
\Psi_h^{(2)}   \approx \!    \frac{- \mu^2} { 4 \pi} \! \! \left( \!  \frac{\tau b^3}{(\tau + b)^5}  \right)^{ \! 1/4} \!  \! \!   \! \!  \sqrt{ \frac{ \tau | \mbit{x}_2 | +  b x_0}{ | \mbit{x}_2 |  }} \exp \! \left( \! -  \frac{( x_0 - | \mbit{x}_2 |)^2}{2(\tau + b)}     \right)  \left(
\begin{array}{c}
\xi_1 \\
 \noalign{\vspace{1ex} }
 \mbit{n}_2 \cdot \mbit{\sigma}_2 \xi_1
\end{array}
\right)   .   \label{Psi^(2)_H Result}
\end{eqnarray}
(Details are relegated to Appendix \ref{AppendixB}.)  This further reduces to
\begin{eqnarray}
\Psi_h^{(2)} 
\approx \!  \frac{- \mu^2} { 4 \pi} \! \left( \!  \frac{\tau b^3}{(\tau + b)^3}  \! \right)^{ \! 1/4} \!  \! \! \!  \exp \! \left( \! - \frac{( x_0 - | \mbit{x}_2 |)^2}{2(\tau + b)}     \right)    \! \left(
\! \! \begin{array}{c}
\xi_1 \\
 \noalign{\vspace{1ex} }
 \mbit{n}_2 \cdot \mbit{\sigma}_2 \xi_1
\end{array} \! \! \! 
\right)   \  ,    \label{2DwithSource phi chi}
\end{eqnarray}
around the peak (\ref{2DmostProbArea}). 

From (\ref{2DfreeResult phi chi}) and (\ref{2DwithSource phi chi}), the total $\Psi^{(2)}$(\ref{generalSolu1}) is given by
\begin{eqnarray}
\Psi^{(2)} \hspace{-4ex} & & = \frac{i \mu^2 a^{1/4} }{4 \sqrt{2 \pi} } \exp \! \left( \! - \frac{(\tau + a + b) }{4a(\tau + b)} ( x_0 - | \mbit{x}_2 |)^2    \right) \! \left(
\! \! \begin{array}{c}
\xi^{(2)} \\
 \noalign{\vspace{1ex} }
 \mbit{n}_2 \cdot \mbit{\sigma}_2 \xi^{(2)}
\end{array} \! \! \! 
\right)   \  ;    \\
\xi^{(2)}  \hspace{-4ex} & & \equiv \exp \! \left( \! - \frac{(\tau - a + b) }{4a(\tau + b)} ( x_0 - | \mbit{x}_2 |)^2    \right) \! \xi_0   \nonumber  \\
& & \hspace{2ex} + i\sqrt{\frac{2}{\pi}}\left( \frac{\tau b^3}{a (\tau + b)^3}   \right)^{\! \! 1/4} \exp \! \left( \!  \frac{(\tau - a + b) }{4a(\tau + b)} ( x_0 - | \mbit{x}_2 |)^2    \right) \! \xi_1 \   . 
\end{eqnarray}
The charge and the current densities are thus found as
\begin{eqnarray}
& & \hspace{-4ex}  \rho^{(2)}_{\rm tot}  \equiv \frac{e}{c} {\Psi^{(2)}}^{\! \dagger} \! \Psi^{(2)} \!    = \frac{2e}{c} \!  \! \left( \frac{\mu^2 a^{1/4} }{4 \sqrt{2 \pi} } \right)^{\! \! 2} \! \!  \exp \! \left( \! - \frac{(\tau + a + b) }{2a(\tau + b)} ( x_0 - | \mbit{x}_2 |)^2    \right)  \! {\xi^{(2)}}^{\! \dagger} \!  \xi^{(2)}    \nonumber   \\
 & & \hspace{-4ex}=   \frac{2e}{c} \!  \! \left( \frac{\mu^2 a^{1/4} }{4 \sqrt{2 \pi} } \right)^{\! \! 2} \! \! \Bigg[ \! \exp \! \left( \! -  \frac{( x_0 - | \mbit{x}_2 |)^2 }{2a}  \right) \!   \xi_0^\dagger \xi_0 \! + \frac{2}{\pi} \! \left( \frac{\tau b^3}{a (\tau + b)^3}   \right)^{ \! \! 1/2}  \! \!  \! \! \! \exp \! \left( \! -  \frac{( x_0 - | \mbit{x}_2 |)^2 }{2(\tau + b) }  \right)  \xi_1^\dagger \xi_1  \nonumber \\
 & & \hspace{6ex} + i\sqrt{\frac{2}{\pi}} \left( \frac{\tau b^3}{a (\tau + b)^3}   \right)^{\! \! 1/4} \! \! \! \! \exp \! \left( \! - \frac{(\tau + a + b) }{2a(\tau + b)} ( x_0 - | \mbit{x}_2 |)^2    \right) \left(   \xi_0^\dagger \xi_1 -  \xi_1^\dagger \xi_0  \right)  \Bigg]    \  , \nonumber  \\
 & &  \hspace{12ex} \mbit{J}^{(2)}_{\rm tot} \equiv e  {\Psi^{(2)}}^\dagger \gamma^0 \mbit{\gamma} \Psi^{(2)}  = \left( c \mbit{n}_2 \rho^{(2)}_{\rm tot}  ,  0 \right)   \  .    \label{2DFullChargeDensity}
\end{eqnarray}

Finally we consider the 1-dimensional case: by restricting $\mbit x$ to be $x_2= x_3=0$ (\ref{sourceDef}) brings (\ref{PsiWithSource}) to 
\begin{eqnarray}
& &  \Psi_h^{(1)}  =  \frac{-\mu^2}{8 \pi^2} \left(  \frac{b }{ \tau (\tau +b )  } \right)^{\! \! 1/4} \! \!  \! \! \exp \! \left( \! - \frac{ x_0^2 }{2 \tau} - \frac{x_1^2}{2 b}  \right) \! \! 
 \left( \! \!
\begin{array}{c} 
\displaystyle{ \tau \partial_0     }  \\
\noalign{\vspace{1ex} }
\displaystyle{  b \sigma_1 \partial_1   }
\end{array} \! \!
\right) \! \!  I_h^{(1)} (x_0, x_1)    \xi_1  \   ,   \label{1DSource phi chi}     \\
& & I_h^{(1)} (x_0, x_1) \equiv \int \frac{d y_1 }{ | y_1| } \exp \left(  - \frac{(\tau +b )}{2 \tau b} y_1^2 - \frac{x_1y_1}{b}  \right) 2 \cosh  \left(  \frac{x_0 |y_1| }{\tau}   \right)   \  .     \label{I_h^(1)}
\end{eqnarray}
After a little calculation (see Appendix \ref{AppendixB}) it reads, when $x_0, x_1 \gg \sqrt{\tau} , \sqrt{b}$,
\begin{eqnarray}
\Psi_h^{(1)} \approx \frac{-\mu^2}{4\sqrt{ 2\pi} }\left(  \frac{ \tau b^3 }{ (\tau +b )^3  } \right)^{\! \! 1/4}  \! \!  \! \! \exp \! \left(-\frac{(x_0-x_1)^2}{2(\tau+b)}\right)   \! \! \left( \! \!
\begin{array}{c}
\xi_1  \\
 \noalign{\vspace{1ex} }
 \sigma_1 \xi_1 
\end{array} \! \!
\right)   .  \label{1DSource phi chi Result}
\end{eqnarray}
From (\ref{1DFree phi chiResult Final}) and (\ref{1DSource phi chi Result}) the total $\Psi^{(1)}$(\ref{generalSolu1}) is obtained  as
\begin{eqnarray}
 \Psi^{(1)} \hspace{-4ex} & & = \frac{ i \mu^2 a^{1/4} }{8 \pi} \!  \exp \! \left( \! - \frac{(\tau + a + b) }{4a(\tau + b)} ( x_0 \! - \! x_1 )^2    \right) \! \! \left(
\! \! \begin{array}{c}
\xi^{(1)} \\
 \noalign{\vspace{1ex} }
 \sigma_1 \xi^{(1)}
\end{array} \! \! \! 
\right)    ;    \\
 \xi^{(1)}  \hspace{-4ex} & & \equiv \exp \! \left( \! - \frac{(\tau - a + b) }{4a(\tau + b)} ( x_0 - x_1)^2    \right) \! \xi_0   \nonumber  \\
& & \hspace{12ex} + i\sqrt{\frac{2}{\pi}} \left( \frac{\tau b^3}{a (\tau + b)^3}   \right)^{\! \! 1/4} \exp \! \left( \!  \frac{(\tau - a + b) }{4a(\tau + b)} ( x_0 - x_1)^2    \right) \! \xi_1 \   . 
\end{eqnarray}
The charge and the current densities are\footnote{Again note that $\mbit{J} = c \mbit{n} \rho$ in (\ref{3DFullChargeDensity}), (\ref{2DFullChargeDensity}), and (\ref{1DFullChargeDensity}) does not imply the current's velocity is $c$. See the footnote \ref{footnote:J=c rho}.} 
\begin{eqnarray}
& & \hspace{-6ex}  \rho^{(1)}_{\rm tot}  \equiv \frac{e}{c} {\Psi^{(1)}}^\dagger  \Psi^{(1)}  = \frac{2e}{c} \left(  \frac{ \mu^2 a^{1/4} }{8 \pi}   \right)^{\! 2} \exp \! \left( \! - \frac{(\tau + a + b) }{2a(\tau + b)} ( x_0 - x_1 )^2    \right)  {\xi^{(1)}}^\dagger  \xi^{(1)}  \nonumber   \\
 & & \hspace{-4ex}=  \!  \frac{2e}{c}  \left(  \frac{ \mu^2 a^{1/4} }{8 \pi}   \right)^{\! 2}  \! \! \Bigg[ \! \exp \! \left( \! -  \frac{( x_0 \!  - \! x_1)^2 }{2a}  \right) \!   \xi_0^\dagger \xi_0 \!  + \frac{\pi}{2} \! \left( \!  \frac{\tau b^3}{a (\tau + b)^3}  \!  \right)^{ \! \! 1/2}  \! \!  \! \! \! \exp \! \left( \! \!  -  \frac{( x_0 \! -  \! x_1)^2 }{2(\tau + b) }  \right)   \! \xi_1^\dagger \xi_1  \nonumber \\
 & & \hspace{6ex} + i\sqrt{\frac{2}{\pi}} \left( \frac{\tau b^3}{a (\tau + b)^3}   \right)^{\! \! 1/4} \! \! \! \! \exp \! \left( \! - \frac{(\tau + a + b) }{2a(\tau + b)} ( x_0 - x_1)^2    \right) \left(   \xi_0^\dagger \xi_1 -  \xi_1^\dagger \xi_0  \right)  \Bigg]    \  ,  \nonumber  \\
 & & \hspace{12ex}   \mbit{J}^{(1)}_{\rm tot} \equiv e  {\Psi^{(1)}}^\dagger  \gamma^0 \mbit{\gamma} \ \Psi^{(1)}   = \left( c  \rho^{(1)}_{\rm tot}  ,  0 ,  0  \right)    .  \label{1DFullChargeDensity}
\end{eqnarray}

\section{Discussion}

In this paper, first we discuss the wave mechanics of $H =c \sqrt{ \mbit{p}^2 + m^2 c^2}$, which tells us that solutions of the Schr\"{o}dinger equation inevitably possess the light-speed portion in the non-relativistic limit $c \mapsto \infty$. The reason is that the kernel contains a derivative acting to a function which owns the discontinuity on the light-cone. The solutions of the Dirac equation are also expressed by differentiations to the invariant functions $\Delta(x)$ and $\Delta_{\rm F}(x)$ which consist of different functions in the time- and the space-like region, thus yield the light-cone singularity, which was the contents of sec.\ref{Sec:The Charge Density of Free Electrons} and \ref{Sec:The Charge and the Current Density for Electrons in a Laboratory}. In relativistic field theories, $c$ appears as $\mu = mc / \hbar$ so that the non-relativistic limit implies $\mu \mapsto \infty$ which, however, also interprets the semiclassical $\hbar \mapsto 0$ or an infinite mass limit $m \mapsto \infty$. According to the last case we can convince ourselves of survival of the light-cone singularity for massive particles. We should emphasize that our conclusion has been derived exclusively in the $x$-representation of $\Psi(x)$ not in the momentum representation.  

The situation is unchanged if the source (\ref{H Definition}) would have a velocity $\mbit{v}$: consider, for example,
\begin{eqnarray}
h^{(3)} (y) \sim  \delta^3 ( \mbit{y} - \mbit{\beta} y_0 )  \exp \! \left( - \frac{y_0^2}{ 2 \tau } \right)     \   ;   \quad  \mbit{\beta} \equiv \frac{ \mbit{v } }{c}      \   . 
\end{eqnarray}
Then from (\ref{PsiWithSource}), 
\begin{eqnarray}
\Psi_h  \sim \int   \!  d^4 y  \delta(x_0-y_0  \mp |\bm x-\bm y|) \delta^3 ( \mbit{y} - \mbit{\beta} y_0 )  \exp \! \left( - \frac{y_0^2}{ 2 \tau } \right)     \nonumber \\
=   \int   \!  d y_0  \delta(x_0-y_0 \mp |\bm x-\mbit{\beta} y_0|)  \exp \! \left( - \frac{y_0^2}{ 2 \tau } \right)    \  . 
\end{eqnarray}
(The spinor part is irrelevant.) Since the zeros in the delta function are given by
$$
x_0-y_0 = \pm |\bm x-\mbit{\beta} y_0| \Longrightarrow y_0 = \left\{ 
\begin{array}{c}
 \left( x_0 + | \mbit{x} | \right)  \left( 1 -  \mbit{\beta } \cdot \mbit{n}  \right)  + O (\beta^2 )        \\
 \noalign{\vspace{1ex}} 
 \left( x_0 - | \mbit{x} | \right)  \left( 1 +   \mbit{\beta } \cdot \mbit{n}  \right)  + O (\beta^2 )   
\end{array}
\right.     \  , 
$$
it reads when $x_0 , \mbit{x} \gg  \sqrt{\tau} $
\begin{eqnarray}
\Psi_h  \sim   \exp \! \left( - \frac{(x_0 - | \mbit{x} |)^2 \left( 1 +   \mbit{\beta } \cdot \mbit{n}  \right)^2  }{ 2 \tau } \right)    \  , 
\end{eqnarray}
which again shows that the maximum signal travels with the speed of light. 

\vspace{1ex}

In order to widen the possibility we finally consider the case that the initial configuration is given with a definite momenta $\mbit{p}$, that is, instead of the packets (\ref{InitialState}) take
\begin{eqnarray}
\psi_{\mbit{p}} (y) \equiv  \frac{1}{ ( a \pi )^{3/4} }  \delta(y_0) \exp \! \left(  \frac{i}{\hbar}   \mbit{p} \cdot \mbit{y}  -  \frac{ \mbit{y}^2 }{ 2 a }   \right) \  \left(\begin{array}{c} \tilde{\xi}_0    \\   
0  \end{array}\right)   \  ,   \qquad    \tilde{\xi}_0^\dagger \tilde{\xi}_0 =1    \  ,     \label{psi_p with p}
\end{eqnarray}
in the solution
\begin{eqnarray}
\Psi_{\mbit{p} } (x) =  \int d^4 y S(x-y) \psi_{\mbit{p}} (y)    \  .    \label{Psi_p(x)}
\end{eqnarray}
Since at $x_0=0$, as was in (\ref{Psi_0 InitialCond}), (\ref{psi_p with p}) implies an initial configuration,
\begin{eqnarray}
\Psi_{\mbit{p} } (x_0=0 ,  \mbit{x} ) = \frac{-i}{ ( a \pi )^{3/4} }  \exp \! \left(  \frac{i}{\hbar}   \mbit{p} \cdot \mbit{x}  -  \frac{ \mbit{x}^2 }{ 2 a }   \right) \left(\begin{array}{c} \tilde{\xi}_0    \\   
0  \end{array}\right)    \  ,  \label{Psi_p InitialCond} 
\end{eqnarray}
whose momentum reads as
\begin{eqnarray}
  \int d^3 \mbit{x} \ \Psi^\dagger_{\mbit{p} } (\mbit{x})  \left( - i \hbar \Nabla \right)  \Psi_{\mbit{p} } (\mbit{x})    = \mbit{p}      \  . 
\end{eqnarray}
Following a similar procedure from (\ref{Psi^(k) Definition}) to (\ref{3DIntegral}), we find
\begin{eqnarray}
\Psi_{\mbit{p} }(x)  = \frac{i\mu^2 }{8\pi ( a \pi)^{3/4} } \exp \! \left(  \frac{i}{\hbar}   \mbit{p} \cdot \mbit{x} - \frac{ x_0^2 + \mbit{x}^2}{2a }   \right)  
\left(
\begin{array}{c} 
1   \\
\noalign{\vspace{1ex} }
\displaystyle{  \frac{a}{x_0}  \mbit{\sigma} \cdot \Nabla  }
\end{array}
\right)    I_{\mbit{p} } \  \tilde{ \xi}_0   \  ,    
\end{eqnarray}
with
\begin{eqnarray}
 I_{\mbit{p} } \equiv \int d^3 \mbit{y} \delta(x_0 -|\bm y|) \exp \! \left(  \frac{i}{\hbar}   \mbit{p} \cdot \mbit{y}  - \frac{ \mbit{x} \cdot \mbit{y} }{a}   \right)  \    ,        \label{I_p Definition}
\end{eqnarray}
which is further rewritten as
\begin{eqnarray}
 I_{\mbit{p} }= \exp \! \left(   - \frac{ia}{\hbar}    \mbit{p}\cdot \Nabla  \right) \int d^3 \mbit{y} \delta(x_0 -|\bm y|)  \exp \! \left(   - \frac{ \mbit{x} \cdot \mbit{y} }{a}   \right)   \hspace{20ex}  \nonumber  \\
   = 2 \pi a x_0  \exp \! \left(   - \frac{ia}{\hbar}    \mbit{p}\cdot \Nabla  \right)\frac{1}{| \mbit{x} | }  \left[  \exp \left(  \frac{| \mbit{x} | x_0 }{ a}  \right)  - \exp \left( - \frac{| \mbit{x} | x_0 }{ a}  \right)  \right]
 \  , 
\end{eqnarray}
with the help of (\ref{3DIntegral Result}).  By noting
\begin{eqnarray}
\Nabla= \mbit{n} \frac{ \partial }{\partial |\mbit{x}| }  \  ;    \qquad  \mbit{n} = \frac{ \mbit{x} }{| \mbit{x} |}   \   ,    \label{Nabla with |x|}
\end{eqnarray}
and that $\displaystyle{ \exp \! \left(   - \frac{ia}{\hbar}  \mbit{p}\cdot \mbit{n} \frac{\partial }{ \partial | \mbit{x} |}  \right) }$ is a shift operator, it reads
\begin{eqnarray}
 I_{\mbit{p} }= 2 \pi a x_0 \frac{\displaystyle{ \exp \!  \left(  \frac{| \mbit{x} | x_0 }{ a} - i \frac{ \mbit{p} \cdot \mbit{n} }{\hbar} x_0 \right) -  \exp \!  \left( - \frac{| \mbit{x} | x_0 }{ a} + i \frac{ \mbit{p} \cdot \mbit{n} }{\hbar} x_0 \right)  } }{| \mbit{x} | - i a \mbit{p} \cdot \mbit{n} / \hbar}    \  .   \label{I_p Result}
\end{eqnarray}
Therefore 
\begin{eqnarray}
& & \hspace{-8ex} \Psi_{\mbit{p} }(x)  = \frac{i\mu^2 a^{1/4}}{4  \pi^{3/4} } \frac{x_0 }{| \mbit{x} | - i a \mbit{p} \cdot \mbit{n} / \hbar}  \nonumber \\
& & \hspace{-4ex}  \times  \left[ \! \!   
\left( \! \! 
\begin{array}{c} 
1   \\
\noalign{\vspace{2ex} }
\displaystyle{ 1 -  \frac{a}{x_0(| \mbit{x} | - i a \mbit{p} \cdot \mbit{n} / \hbar) }  }
\end{array} \! \! 
\right) \! \exp \! \left( - \frac{i (\mbit{p} \cdot \mbit{n} x_0 -  \mbit{p} \cdot \mbit{x} ) }{\hbar}  - \frac{(x_0 - |\mbit{x} | )^2}{2a}   \right)  \right.  \nonumber   \\
& &  \hspace{-4ex} \left.   +  \! \! \left( \! \! \! \! 
\begin{array}{c} 
-1   \\
\noalign{\vspace{2ex} }
\displaystyle{ 1 +  \frac{a}{x_0(| \mbit{x} | - i a \mbit{p} \cdot \mbit{n} / \hbar) }  }
\end{array} \! \! \!
\right) \! \exp \! \left(  \frac{i (\mbit{p} \cdot \mbit{n} x_0 -  \mbit{p} \cdot \mbit{x} ) }{\hbar}  - \frac{(x_0 + |\mbit{x} | )^2}{2a}   \right) \! \! \right] \! \! \! \left( \! \! \!
\begin{array}{c} 
 \tilde{\xi}_0    \\
\noalign{\vspace{2ex} }
  \mbit{\sigma} \cdot \mbit{n}  \tilde{\xi}_0  
\end{array} \! \! \!  
\right)   . 
\end{eqnarray}
When $x_0 , | \mbit{x}| \gg  \sqrt{a} \ ;  \  | \mbit{x} | - i a \mbit{p} \cdot \mbit{n} / \hbar \approx | \mbit{x} |$,  it yields to a plane wave, 
\begin{eqnarray}
 \Psi_{\mbit{p} }(x) \approx \frac{i\mu^2 a^{1/4}}{4  \pi^{3/4} } \exp \! \left( - \frac{i (\mbit{p} \cdot \mbit{n} x_0 -  \mbit{p} \cdot \mbit{x} ) }{\hbar}  \right)  \left( \! \! \!
\begin{array}{c} 
 \tilde{\xi}_0    \\
\noalign{\vspace{2ex} }
  \mbit{\sigma} \cdot \mbit{n}  \tilde{\xi}_0  
\end{array} \! \! \!  
\right)  \  , 
\end{eqnarray}
around the peak $x_0 - |\mbit{x}| \approx 0$, which implies that the energy-momentum relation is given by
\begin{eqnarray}
E = c| \mbit{p} |    \  .    
\end{eqnarray}
Therefore an alternative way to observe the relativistic remnant is a measurement for the energy and the momentum of electrons in the vacuum.

\vspace{4ex}

\noindent{\Large \bf Acknowledgements}

\vspace{2ex}

The authors are grateful to Hiroto So for discussions.  T. K. also thanks to Tadashi Toyoda for useful comments. 

\appendix

\section{The Derivation of  (\ref{H^(2)-K zero})}\label{AppendixA}

Owing to the delta function $\delta (ct -r)$
$$
\chi_{\pm} = \mu \sqrt{ \pm \left(  (ct)^2  -r^2 \right) } \mapsto 0  \  , 
$$
so that
\begin{eqnarray}
 & & \hspace{-6ex}  \frac{\pi }{2}   \frac{   H_1^{(2)} \! \left( \chi_{+} \right) }{\chi_{+}} \Big|_{\chi_{+}=0} \hspace{-1ex} + i  \frac{  K_1 \! \left( \chi_{-} \right) }{\chi_{-}} \Big|_{\chi_-=0}    \nonumber  \\ 
& &   =  \frac{\pi }{2}  \frac{   J_1  \! \left( \chi_{+} \right) }{\chi_{+}}\Big|_{\chi_{+}=0}  
\hspace{-1ex}  - i \left\{   \frac{\pi }{2} \frac{   N_1 \! \left( \chi_{+} \right) }{\chi_{+}}\Big|_{\chi_{+}=0} \hspace{-1ex}  -  \frac{  K_1 \! \left( \chi_{-} \right) }{\chi_{-}} \Big|_{\chi_-=0} \right\}  \ ,   \label{H_1-K_1Term}
\end{eqnarray}
where we have noticed
\begin{eqnarray}
H^{(2)}_n (z)= J_n (z) -iN_n (z) \ , \qquad    (n=0,1,2, \dots )  .    \label{H^(2) Definition}
\end{eqnarray}
From the definition of $J_n$ and $I_n$\cite{rf:Ryz5},
\begin{eqnarray}
& &  J_n (z) = \left( \frac{z}{2} \right)^n  \sum_{k=0}^\infty \frac{(-)^k }{k! (n + k)! }\left( \frac{z}{2} \right)^{2k}   \ ,   \label{Def Jnu} \\
& &  I_n (z) = \left( \frac{z}{2} \right)^n  \sum_{k=0}^\infty \frac{1}{k! (n + k)! }\left( \frac{z}{2} \right)^{2k}    \ ,    \label{Def Inu} 
\end{eqnarray}
we have
\begin{equation}
\frac{   J_1  \! \left( \chi_{+} \right) }{\chi_{+}} \Big|_{\chi_{+}=0}= \frac{1}{2}  \ ; \quad \frac{   I_1  \! \left( \chi_{-} \right) }{\chi_{-}} \Big|_{\chi_{-}=0}= \frac{1}{2}  \   .     \label{J_1/z I_1/z at z=0}
\end{equation}
While 
\begin{eqnarray*}
  \frac{\pi }{2} \frac{   N_1 \! \left( \chi_{+} \right) }{\chi_{+}}\Big|_{\chi_{+}=0}   & \hspace{-3ex} = \hspace{-1ex}&   \frac{   J_1  \! \left( \chi_{+} \right) }{\chi_{+}} \left( \gamma + \ln \frac{\chi_{+}}{2} \right)   \Big|_{\chi_{+}=0} \hspace{-1ex} -\frac{1}{4} - \frac{1}{\chi_{+}^2} \Big|_{\chi_{+}=0} \ ,   \\ 
  \frac{  K_1 \! \left( \chi_{-} \right) }{\chi_{-}} \Big|_{\chi_-=0}& \hspace{-3ex} = \hspace{-1ex} &  \frac{   I_1  \! \left( \chi_{-} \right) }{\chi_{-}} \left( \gamma + \ln \frac{\chi_{-}}{2} \right)   \Big|_{\chi_{-}=0} \hspace{-1ex} + \frac{1}{4} + \frac{1}{\chi_{-}^2} \Big|_{\chi_{-}=0} \ , 
\end{eqnarray*}
with Euler's constant $\gamma$, in view of \cite{rf:Ryz7}
\begin{eqnarray}
& & \hspace{-8ex}\frac{\pi }{2}N_n(z) = J_n (z) \left( \gamma + \ln \frac{z}{2} \right)  \nonumber \\
& & \hspace{3ex}-\frac{1}{2} \left( \frac{z}{2} \right)^n \sum_{k=0}^\infty \frac{(-)^k }{k! (n+k)! }\left( \frac{z}{2} \right)^{2k} \left[ \sum_{m=1}^k \frac{1}{m}+ \sum_{m=1}^{n+k} \frac{1}{m}\right]  \nonumber \\
& & \hspace{3ex} -\frac{1}{2} \left( \frac{2}{z} \right)^n \sum_{k=0}^{n-1} \frac{(n-k-1)!}{k!} \left( \frac{z}{2} \right)^{2k}  \ ; \label{N_n Def}  \\
& & \hspace{-8ex} K_n(z) = (-)^{n+1}I_n (z) \left( \gamma + \ln \frac{z}{2} \right)  \nonumber \\
& &\hspace{3ex} -\frac{(-)^n}{2} \left( \frac{z}{2} \right)^n \sum_{k=0}^\infty \frac{(-)^k }{k! (n+k)! }\left( \frac{z}{2} \right)^{2k} \left[ \sum_{m=1}^k \frac{1}{m}+ \sum_{m=1}^{n+k} \frac{1}{m}\right]  \nonumber \\
& &\hspace{3ex} + \frac{1}{2} \left( \frac{2}{z} \right)^n \sum_{k=0}^{n-1} \frac{(n-k-1)!}{k!} \left( \frac{z}{2} \right)^{2k} \  , \label{K_n Def} 
\end{eqnarray}
then
\begin{eqnarray}
& & \hspace{-4ex}  \frac{\pi }{2} \frac{   N_1 \! \left( \chi_{+} \right) }{\chi_{+}}\Big|_{\chi_{+}=0} - \hspace{0ex}  \frac{  K_1 \! \left( \chi_{-} \right) }{\chi_{-}} \Big|_{\chi_-=0}  \nonumber   \\
& & =  \frac{1}{2}\left(  \ln \sqrt{x_\mu^2}-\ln \sqrt{-x^2_\mu }  \right) \Bigg|_{x_\mu^2=0} \hspace{-2ex} -\frac{1}{2} - \left( \frac{1}{\mu^2 x_\mu^2}  - \frac{1}{\mu^2 x_\mu^2} \right) \Bigg|_{x_\mu^2=0}   \nonumber \\ 
& &  \hspace{-2ex} =  \frac{1}{4}  \Big[  \ln (ct -r) + \ln (ct+r)   - \ln (r-ct) -\ln (r+ct)  \Big] \Bigg|_{ct=r} \hspace{-2ex} - \frac{1}{2} =  -\frac{i \pi }{4}  - \frac{1}{2}  \ ,    \label{N_1/z -K_1/z}
\end{eqnarray}
where we have used $\ln (r-ct) = i \pi +  \ln( ct -r) $ to the final expression.  

Therefore (\ref{H_1-K_1Term}) reads
\begin{eqnarray}
 \frac{\pi }{2}   \frac{   H_1^{(2)} \! \left( \chi_{+} \right) }{\chi_{+}} \Big|_{\chi_{+}=0} \hspace{-1ex} + i  \frac{  K_1 \! \left( \chi_{-} \right) }{\chi_{-}} \Big|_{\chi_-=0} = \frac{\pi}{4} -  \frac{\pi}{4} + \frac{i}{2}  =  \frac{i}{2} \   .     \quad   \dbox
\end{eqnarray}

\section{Derivations of $\Psi_h^{(3)}, \Psi_h^{(2)}$ and $\Psi_h^{(1)}$.}\label{AppendixB}

(\ref{I_N^(3) Result}):  With the aid of the polar coordinates, (\ref{I_N^(3)}) becomes
\begin{eqnarray}
I_h^{(3)} (x_0, \mbit{x} ) \hspace{-4ex} & & = \frac{8 \pi b}{ | \mbit{x} | } \! \! \int_{0}^{\infty} \! \! dy \exp \! \left( - \frac{(\tau + b) y^2}{2 \tau b}  \right) \cosh \left(  \frac{x_0  }{ \tau}  y \right)  \sinh \left( \frac{ | \mbit{x} |}{ b} y \right)   \nonumber \\
& & \hspace{-6ex} = \frac{4 \pi b}{ | \mbit{x} | } \! \! \int_{0}^{\infty}  \! \! dy \exp \! \left( - \frac{(\tau + b) y^2}{2 \tau b}  \right) \! \!  \left[  \sinh \left( \!  \frac{bx_0 + \tau | \mbit{x} | }{ \tau b}  y  \right) \! -  \sinh \left( \!  \frac{bx_0 - \tau | \mbit{x} | }{ \tau b} y \!   \right)  \! \right]   ,   \label{I_h^(3)-1}
\end{eqnarray}
where the addition theorem for the hyperbolic function has been used. In view of the error function formula (\ref{errorFunc Def}) we have
\begin{eqnarray}
 \int_{0}^{\infty}  \! \! dy {\mathrm e}^{- A y^2} \sinh B y = \frac{1}{2} \sqrt{\frac{\pi }{A}} \mathrm{erf} \! \left(  \frac{B}{ 2 \sqrt{A} }  \right) \exp \! \left( \frac{B^2}{4A}  \right)   \  ,  \label{errorFnFormula}
\end{eqnarray}
so that (\ref{I_h^(3)-1}) becomes (\ref{I_N^(3) Result}).

\vspace{2ex}

\noindent (\ref{3Dint Psi}) $\Psi_h^{(3)}$ : In view of (\ref{I_N^(3) Result}) and (\ref{3D withH phi chi Definition}) note
\begin{eqnarray}
\exp \! \left( - \frac{x_0^2}{ 2 \tau} -  \frac{\mbit{x}^2}{ 2 b} +   \frac{( b x_0 \pm \tau| \mbit{x} | )^2}{2 \tau b (\tau + b)}        \right) = \exp \! \left( - \frac{ (x_0 \mp | \mbit{x} |)^2 }{ 2 (\tau +b) }    \right)  \ ,   \label{To Lightcone}
\end{eqnarray}
then apply $ \partial_0$ and $ \Nabla $ to obtain
\begin{eqnarray*}
& & \hspace{-5ex}\varphi_h^{(3)}  (x)   = \frac{-\mu^2}{ 2 \sqrt{2 \pi} } \left(  \frac{\tau b^3 }{(\tau +b )^7} \right)^{\! \! 1/4} \! \! \left[ \frac{bx_0 + \tau | \mbit{x} | }{| \mbit{x} |}  {\rm erf} \!  \left(  \frac{ b x_0 + \tau| \mbit{x} |}{ \sqrt{2 \tau b (\tau + b) } } \right)  \!  \exp \! \left(  - \frac{(  x_0 - | \mbit{x} | )^2}{2 (\tau + b)}   \right)  \right. \nonumber  \\
& & \hspace{8ex}\left.   - \frac{bx_0 - \tau | \mbit{x} | }{| \mbit{x} |}  {\rm erf} \!  \left(  \frac{ b x_0 - \tau| \mbit{x} |}{ \sqrt{2 \tau b (\tau + b) } } \right)  \!  \exp \! \left(  - \frac{(  x_0 + | \mbit{x} | )^2}{2 (\tau + b)}   \right)     \right]  \ ,  \\
& &  \hspace{-5ex} \chi_h^{(3)}  (x)   = \frac{-\mu^2}{ 2 \sqrt{2 \pi} }  \left(  \frac{\tau b^3 }{(\tau +b )^7} \right)^{ \! \! 1/4} \! \!  \nonumber  \\
& &  \hspace{8ex} \times  \left[ \left(  \frac{bx_0 + \tau | \mbit{x} | }{| \mbit{x} |} -  \frac{b (\tau + b) }{ \mbit{x}^2 }    \right) {\rm erf} \!  \left(  \frac{ b x_0 + \tau| \mbit{x} |}{ \sqrt{2 \tau b (\tau + b) } } \right)  \!  \exp \! \left(  - \frac{(  x_0 - | \mbit{x} | )^2}{2 (\tau + b)}   \right)  \right. \nonumber  \\
 & & \hspace{8ex}  + \left(  \frac{bx_0 - \tau | \mbit{x} | }{| \mbit{x} |}  +   \frac{b (\tau + b) }{ \mbit{x}^2 }  \right){\rm erf} \!  \left(  \frac{ b x_0 - \tau| \mbit{x} |}{ \sqrt{2 \tau b (\tau + b) } } \right)  \!  \exp \! \left(  - \frac{(  x_0 + | \mbit{x} | )^2}{2 (\tau + b)}   \right)    \nonumber  \\ 
 & &  \hspace{8ex} \left.   + \sqrt{ \frac{2 \tau b (\tau + b)}{ \pi } } \frac{2}{| \mbit{x} |}  \exp \! \left(  - \frac{x_0^2}{ 2 \tau} - \frac{ \mbit{x}^2}{2 b}  \right)   \right] (\mbit{n} \cdot  \mbit{\sigma} ) \xi_1   \ . 
\end{eqnarray*}
When $x_0, |\mbit{x}| \gg \sqrt{\tau} , \sqrt{b}$, terms vanish, except the first one in the right hand side to give 
\begin{eqnarray*}
\Psi_h^{(3)} \approx  \frac{-\mu^2}{ 2 \sqrt{2 \pi} } \left(  \frac{\tau b^3 }{(\tau +b )^7} \right)^{\! \! 1/4}  \frac{bx_0 + \tau | \mbit{x} | }{| \mbit{x} |}  \hspace{28ex}   \nonumber \\
\times  {\rm erf} \!  \left(  \frac{ b x_0 + \tau| \mbit{x} |}{ \sqrt{2 \tau b (\tau + b) } } \right)  \!  \exp \! \left(  - \frac{(  x_0 - | \mbit{x} | )^2}{2 (\tau + b)}   \right) \left(
\begin{array}{c}
  \xi_1   \\
  \mbit{n} \cdot \mbit{ \sigma} \xi_1 
 \end{array}
\right)   \  ,  
\end{eqnarray*}
which further, by noting
$$
\lim_{ X \mapsto \infty} {\rm erf} (X) =1   \   ;    \qquad   X \equiv \frac{ b x_0 + \tau| \mbit{x} |}{ \sqrt{2 \tau b (\tau + b) } } \   , 
$$
becomes
\begin{eqnarray*}
\Psi_h^{(3)} \approx \frac{-\mu^2}{ 2 \sqrt{2 \pi} }  \left(  \frac{\tau b^3 }{(\tau +b )^7} \right)^{\! \! 1/4}  \frac{bx_0 + \tau | \mbit{x} | }{| \mbit{x} |}  \exp \! \left(  - \frac{(  x_0 - | \mbit{x} | )^2}{2 (\tau + b)}   \right) \left(
\begin{array}{c}
  \xi_1   \\
  \mbit{n} \cdot \mbit{ \sigma} \xi_1 
 \end{array}
\right)   \  ,    
\end{eqnarray*}
yielding to (\ref{3Dint Psi}) around the peak (\ref{mostProbabArea}).

\vspace{2ex}

\noindent (\ref{I^(2)_H Result}): (\ref{2DInt withSouce}) reads by use of the polar coordinate
\begin{eqnarray*}
I_h^{(2)} (x_0, \mbit{x}_2 )= 2 \int_{0}^{\infty} dy \exp \left(  - \frac{(\tau +b )}{2 \tau b} y^2 \right)\cosh \left(  \frac{x_0 y}{ \tau}  \right)   \int_{0}^{2 \pi}  d \theta \exp \left(  - \frac{ | \mbit{x}_2 |y }{b} \cos \theta \right)  \   .  
\end{eqnarray*}
When $ | \mbit{x}_2 | \gg \sqrt{b}$, the saddle point method around $\theta = \pi$ brings the angular part to
$$
\int_{0}^{2 \pi}  d \theta \exp \left(  - \frac{ | \mbit{x}_2 |y }{b} \cos \theta \right) = \sqrt{ \frac{2 \pi b}{| \mbit{x}_2 |y}   }  \exp \left(   \frac{ | \mbit{x}_2 |y }{b}  \right) \left( 1 + O \! \left( \frac{1}{| \mbit{x}_2 |}   \right)  \right) \  . 
$$
Then 
\begin{eqnarray*}
I_h^{(2)} (x_0, \mbit{x}_2 ) \approx \sqrt{ \frac{2 \pi b}{| \mbit{x}_2 |}   } \int_{0}^{\infty} \frac{ dy}{\sqrt{y}}  \Bigg[     \exp   \left(  - \frac{(\tau +b )}{2 \tau b} y^2 +    \frac{(\tau | \mbit{x}_2 |  +b x_0 )}{ \tau b} y  \right)  \\
 +  \exp  \left(  - \frac{(\tau +b )}{2 \tau b} y^2 +    \frac{(\tau | \mbit{x}_2 |  - b x_0 )}{ \tau b} y  \right)  \Bigg]    \   , 
\end{eqnarray*}
which is further rewritten, in terms of a dimensionless quantity $Y$,
$$
  y = \frac{ \tau | \mbit{x}_2 |  \pm b x_0 }{ \sqrt{\tau b}}Y  \   , 
$$ 
as
\begin{eqnarray}
I_h^{(2)} (x_0, \mbit{x}_2 ) =  \sqrt{ \frac{2 \pi b}{| \mbit{x}_2 |}   } \left(  F^{(+)} + F^{(-)}  \right)  \  ,   \hspace{20ex}   \label{I_h^(2) by F+ F-}   \\
F^{(\pm)} \equiv \sqrt{\frac{\tau | \mbit{x}_2 |  \pm b x_0}{\sqrt{\tau b}  }} \int_{0}^{\infty} \frac{ dY}{\sqrt{Y}}    \exp   \left(  - \frac{(\tau +b ) (\tau | \mbit{x}_2 |  \pm b x_0 )^2}{( \tau b)^2}  \left\{  \frac{ Y^2 }{2} -     \frac{\sqrt{\tau b} }{ \tau +b } Y   \right\} \right)   \   .  \nonumber
\end{eqnarray}
Since $| \mbit{x}_2 |, x_0 \gg \sqrt{\tau}, \sqrt{b}$, the saddle point method around $Y= \sqrt{\tau b} /(\tau + b) $ gives
\begin{eqnarray*}
F^{(\pm)} = \sqrt{\frac{2 \pi \tau b }{\tau | \mbit{x}_2 |  \pm b x_0}} \exp \left(  \frac{( \tau | \mbit{x}_2 |  \pm b x_0 )^2}{2 \tau b (\tau + b) }   \right) \left[   1 + O \! \left( \frac{1}{( \tau | \mbit{x}_2 |  \pm b x_0 )^2}       \right)   \right]    \  . 
\end{eqnarray*}
Inserting this into (\ref{I_h^(2) by F+ F-}) we have (\ref{I^(2)_H Result}).

\vspace{3ex}

\noindent (\ref{Psi^(2)_H Result}) $\Psi_h^{(2)}$: Apply differentiations in (\ref{2D phi chi}) and note the relation (\ref{To Lightcone}) by putting $\mbit{x} \mapsto \mbit{x}_2$ to obtain
\begin{eqnarray}
\Psi_h^{(2)} \approx  \frac{- \mu^2} { 4 \pi} \left(  \frac{\tau b^3}{(\tau + b)^5}  \right)^{1/4} \Bigg[  \sqrt{ \frac{ \tau | \mbit{x}_2 | +  b x_0}{ | \mbit{x}_2 |  }} \exp \left(- \frac{( x_0 - | \mbit{x}_2 |)^2}{2(\tau + b)}     \right) \nonumber   \\
\mp   \sqrt{ \frac{ \tau | \mbit{x}_2 | -  b x_0}{ | \mbit{x}_2 |  }} \exp \left(- \frac{( x_0 + | \mbit{x}_2 |)^2}{2(\tau + b)}     \right)  \Bigg]  \left(
\begin{array}{c}
\xi_1 \\
 \noalign{\vspace{1ex} }
 \mbit{n}_2 \cdot \mbit{\sigma}_2 \xi_1
\end{array}
\right)  \   ,    \label{Psi^(2)_H -1}
\end{eqnarray}
where we have omitted terms of 
$$
O \! \left( \frac{\tau b (\tau + b) }{( \tau | \mbit{x}_2 | \pm bx_0)^2}\right)  \ , \qquad O \! \left( \frac{ b (\tau + b) }{| \mbit{x}_2 | ( \tau | \mbit{x}_2 | \pm bx_0)}\right)  \  . 
$$
Under $x_0, |\mbit{x}_2| \gg \sqrt{\tau}, \sqrt{b}$, (\ref{Psi^(2)_H -1}) becomes (\ref{Psi^(2)_H Result}).

\vspace{3ex}

\noindent (\ref{1DSource phi chi Result}) $\Psi_h^{(1)}$: (\ref{I_h^(1)}) becomes (we have put $y_1 \mapsto y$)
\begin{eqnarray}
I_h^{(1)} (x_0, x_1) = 2 \! \!  \int_{0}^{\infty} \!  \frac{d y }{ y } \exp \!  \left( \!  - \frac{(\tau +b )}{2 \tau b} y^2 \!  \right) \!  \left[ \cosh \!  \left(  \frac{bx_0 + \tau x_1 }{\tau b}  y  \! \right) \!  + \cosh  \!  \left(  \frac{bx_0 - \tau x_1 }{\tau b}  y \!  \right) \!   \right]    .  
\end{eqnarray}
Apply the differentiation in (\ref{1DSource phi chi}), with the aid of the error function formula (\ref{errorFnFormula}), to find
\begin{eqnarray}
 \Psi_h^{(1)} \! (x_0, x_1) =  \frac{-\mu^2}{4\sqrt{ 2\pi} }\left(  \frac{ \tau b^3 }{  (\tau +b )^3  } \right)^{\! \! 1/4}   \Bigg[  {\rm erf}\left( \frac{bx_0 + \tau x_1 }{2 \tau b (\tau + b)} \right)    \exp \! \left(-\frac{(x_0-x_1)^2}{2(\tau+b)}\right)  \nonumber \\
  \pm  {\rm erf}\left( \frac{bx_0 - \tau x_1 }{2 \tau b (\tau + b)} \right)  \exp \! \left(-\frac{(x_0+x_1)^2}{2(\tau+b)}\right) \! \Bigg]  \left( \! \!
\begin{array}{c}
  \xi_1  \\
 \noalign{\vspace{1ex} }
 \sigma_1 \xi_1 
\end{array} \! \!
\right)    \   ,  
\end{eqnarray}
which yields, when $x_0, x_1 \gg \sqrt{\tau} , \sqrt{b}$, to (\ref{1DSource phi chi Result}).



%

\begin{thebibliography}{99}
\bibitem{rf:Ryz6} p.880, 8.250-1 in the reference\cite{rf:Ryz}. (Instead of ${\rm \Phi}(x)$ we write ${\rm erf}(x)$.)
\end{thebibliography}
\end{document}